\documentclass[10pt,journal,final,twocolumn]{IEEEtran}

\usepackage{graphicx}
\usepackage{subfigure}
\usepackage{epstopdf}
\usepackage{epsfig}
\usepackage[cmex10]{amsmath}
\usepackage{amssymb}
\usepackage{times}
\usepackage{ntheorem}
\usepackage{pifont}
\usepackage{stfloats}
\usepackage{bm}
\usepackage{color}
\usepackage{makecell}
\usepackage{array}
\usepackage{algorithm}
\usepackage{algorithmic}
\usepackage{multirow}

\begin{document}
%
\title{Framework on Deep Learning Based Joint Hybrid Processing for mmWave Massive MIMO Systems}

\author{\IEEEauthorblockN{Peihao~Dong,~\IEEEmembership{Member,~IEEE}, Hua Zhang,~\IEEEmembership{Member,~IEEE},
and Geoffrey Ye Li,~\IEEEmembership{Fellow,~IEEE}
}

\thanks{

P. Dong and H. Zhang are with the National Mobile Communications Research Laboratory, Southeast University, Nanjing 210096, China (e-mail: phdong@seu.edu.cn; huazhang@seu.edu.cn).

G. Y. Li is with the School of Electrical and Computer Engineering, Georgia Institute of Technology, Atlanta, GA 30332 USA (e-mail: liye@ece.gatech.edu).
}
}

\IEEEtitleabstractindextext{%
\begin{abstract}
For millimeter wave (mmWave) massive multiple-input multiple-output (MIMO) systems, hybrid processing architecture is essential to significantly reduce the complexity and cost but is quite challenging to be jointly optimized over the transmitter and receiver. In this paper, deep learning (DL) is applied to design a novel joint hybrid processing framework (JHPF) that allows end-to-end optimization by using back propagation. The proposed framework includes three parts: hybrid processing designer, signal flow simulator, and signal demodulator, which outputs the hybrid processing matrices for the transceiver by using neural networks (NNs), simulates the signal transmission over the air, and maps the detected symbols to the original bits by using the NN, respectively. By minimizing the cross-entropy loss between the recovered and original bits, the proposed framework optimizes the analog and digital processing matrices at the transceiver jointly and implicitly instead of approximating pre-designed label matrices, and its trainability is proved theoretically. It can be also directly applied to orthogonal frequency division multiplexing systems by simply modifying the structure of the training data. Simulation results show the proposed DL-JHPF outperforms the existing hybrid processing schemes and is robust to the mismatched channel state information and channel scenarios with the significantly reduced runtime.
\end{abstract}

}

\maketitle

\IEEEdisplaynontitleabstractindextext

%
\IEEEpeerreviewmaketitle

\section{Introduction}

Due to the huge bandwidth, millimeter wave (mmWave) communications have been recognized as one of the key technologies to meet the demand for unprecedentedly high data rate transmission in the future mobile networks \cite{A. L. Swindlehurst}. By equipping large-scale antenna arrays, massive multiple-input multiple-output (MIMO) can provide sufficiently large array gains for spatial multiplexing and beamforming \cite{L. Lu}. MmWave massive MIMO communications can obtain the merits of both of them and thus have attracted significant interest \cite{B. Wang}. However, the expensive and power-hungry hardwares used in mmWave bands become the main obstacle to equipping a dedicated radio frequency (RF) chain for each antenna. The mainstream solution for this problem is to use the two-stage hybrid architecture, where a large number of antennas are connected to much fewer RF chains via phase shifters \cite{L. Liang}, \cite{L. Pan}.

\subsection{Related Work}

For mmWave massive MIMO systems with the hybrid architecture, both the analog and digital processing should be carefully designed to achieve the comparable performance to the fully-digital systems. In \cite{L. Liang}, a low-complexity hybrid precoding scheme at the base station (BS) has been proposed for the massive MIMO downlink with single-antenna users. The hybrid architecture has been further introduced to the user side in \cite{W. Ni}, where hybrid block diagonalization (HBD) has been used for the analog and digital processing design. By exploiting the sparsity of mmWave channels, the hybrid precoding and combining at both the transmitter and receiver have been optimized in \cite{O. E. Ayach}. The heuristic hybrid beamforming design in \cite{F. Sohrabi} can approach the performance of the fully-digital architecture. The alternating minimization algorithms for both fully-connected and sub-connected hybrid architectures in \cite{X. Yu} are with low complexity and limited performance loss. In \cite{L. Zhao}, the hybrid processing along with channel estimation has been designed and analyzed for both the sparse and non-sparse channels. The uniform channel decomposition and nonlinear digital processing have been introduced in \cite{Y. Lin} for hybrid beamforming design. In the existing works, the hybrid processing matrices at the transmitter and receiver are usually optimized separately due to the intractability of the joint optimization with non-convex constraints, which makes the further performance improvement possible with joint optimization.

Deep learning (DL) has achieved great success in various fields, including computer vision \cite{K. He_a}, speech signal processing \cite{A. Graves}, natural language processing \cite{R. Collobert}, and so on, due to its unique ability in extracting and learning inherent features. It has been recently introduced to wireless communications and shown quite powerful in the optimization of communication systems \cite{T. O'hea}--\cite{G. Gui} and resource allocation \cite{L. Liang_b}--\cite{Z. Yang}. In \cite{H. Ye_a}, DL has been successfully applied in pilot-assisted signal detection for orthogonal frequency division multiplexing (OFDM) systems with non-ideal transceiver and channel conditions. For wideband mmWave massive MIMO systems in time-varying channels, channel correlation has been exploited by deep convolutional neural network (CNN) in \cite{P. Dong_b} to improve the accuracy and accelerate the computation for the channel estimation. Deep neural network (DNN) has been utilized in \cite{S. Gao} to model the mapping relationship among antennas for reliable channel estimation in massive MIMO systems with mixed-resolution ADCs. An autoencoder-like DNN has been developed in \cite{C.-K. Wen} to reduce the overhead for channel state information (CSI) feedback in the frequency duplex division massive MIMO system. In \cite{C. Lu_a}, CNN has been utilized in CSI compression and uncompression to significantly improve the recovery accuracy. By combining the residual network and CNN, an efficient channel quantization scheme has been proposed from the perspective of bit-level in \cite{C. Lu_b}. The DL based end-to-end optimization has been developed in \cite{S. Dorner} and \cite{H. Ye_b} by breaking the block structures at the transceiver. DL has been recently used to design the hybrid processing matrices for massive MIMO systems with various transceiver architectures \cite{H. Huang}--\cite{X. Bao}. In \cite{H. Huang}, the analog and digital precoder design has been modeled as the DNN mapping based on geometric mean decomposition. In \cite{T. Lin}, DNN has been applied to design the analog precoder for massive multiple-input single-output (MISO) systems. Deep CNN has been applied to learn the phases of the analog precoder and combiner for mmWave massive MIMO systems in \cite{A. M. Elbir}. For the same system, channel estimation and analog processing have been jointly optimized by DL with reduced pilot overhead in \cite{X. Li}. In \cite{X. Bao}, deep CNN along with an equivalent channel hybrid precoding algorithm have been proposed to design the hybrid processing matrices.

\subsection{Motivation and Contribution}

The research on the DL based hybrid processing for mmWave massive MIMO systems is still in the exploratory stage and has many open issues. The existing works have applied DL to design the analog precoder \cite{T. Lin}, the analog combiner \cite{X. Bao}, the analog precoder and combiner \cite{A. M. Elbir}, \cite{X. Li}, and the analog and digital precoders \cite{H. Huang}. Currently, only partial hybrid processing is designed by DL for the mmWave transceiver. In addition, conventional hybrid processing schemes are usually used to generate label matrices for the DNN to approximate, which limits the performance of the DL based approaches. The problems in the existing works motivate us to propose a general DL based joint hybrid processing framework (DL-JHPF) with the following two unique features:
\begin{itemize}[\IEEEsetlabelwidth{Z}]
\item[1)] The framework jointly optimizes the analog and digital processing matrices at both the transmitter and receiver in an end-to-end manner without pre-designed label matrices. By doing this, it can be applied to various types of mmWave transceiver architectures and will have the potential to break through the performance of the existing schemes.

\item[2)] The framework enables end-to-end optimization but still preserves the block structures at the transceiver considering the hardware and power constraints in practical implementation for the hybrid architecture, which is quite different from the end-to-end optimization in \cite{S. Dorner} and \cite{H. Ye_b}.
\end{itemize}
The main contributions of this paper are summarized as follows.
\begin{itemize}[\IEEEsetlabelwidth{Z}]
\item[1)] We model the joint analog and digital processing design for the transceiver as a DL based framework, which consists of the NN based hybrid processing designer, signal flow simulator, and NN based signal demodulator. For the sake of practical implementation, it does not break the original block structures at the transceiver but still allows the back-propagation (BP) based end-to-end optimization by minimizing the cross-entropy loss between recovered and original bits. The trainability of DL-JHPF is proved theoretically.

\item[2)] We extend the proposed framework to OFDM systems by simply modifying the structure of the training data. The extension does not complicate the framework architecture and guarantees the relatively short training time even if the number of subcarriers is large.

\item[3)] We verify the effectiveness of the proposed framework by numerical results based on the 3rd Generation Partnership Project (3GPP) channel model that can well depict the real channel environment. The proposed DL-JHPF achieves remarkable improvement in bit-error rate (BER) performance even with mismatched CSI and channel scenarios. Thanks to the careful design, DL-JHPF reduces the runtime significantly by sufficiently exploiting the parallel computing and thus is more suitable for rapidly varying mmWave channels.
\end{itemize}

The rest of the paper is organized as follows. Section II describes the channel model and signal transmission process for the considered mmWave massive MIMO system. The proposed DL-JHPF is elaborated in Section III. Simulation results are provided in Section IV to verify the effectiveness of the proposed framework and finally Section V gives concluding remarks.

\emph{Notations}: In this paper, we use upper and lower case boldface letters to denote matrices and vectors, respectively. $\lVert\cdot\rVert_{F}$, $(\cdot)^T$, $(\cdot)^H$, and $\mathbb{E}\{\cdot\}$ represent the Frobenius norm, transpose, conjugate transpose, and expectation, respectively. $\mathcal{CN}(\mu,\sigma^2)$ represents circular symmetric complex Gaussian distribution with mean $\mu$ and variance $\sigma^2$. $[\mathbf{X}]_{i,j}$ and $[\mathbf{x}]_{i}$ denote the $(i,j)$th element of matrix $\mathbf{X}$ and the $i$th element of vector $\mathbf{x}$, respectively. $|\cdot|$ denotes the amplitude of a complex number.

\section{System Model}

\begin{figure}[t]
\centering
\includegraphics[width=3.5in]{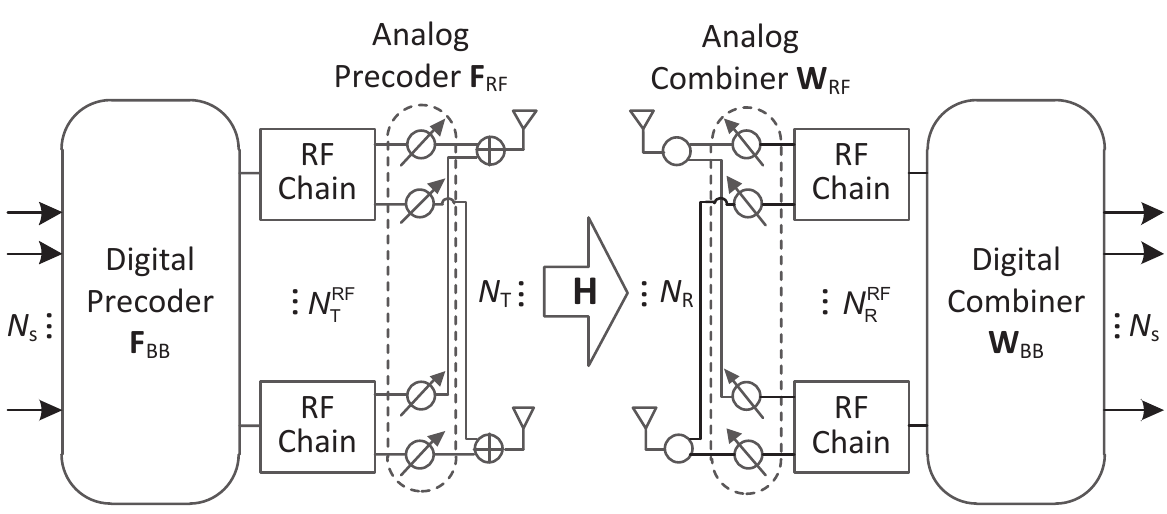}
\caption{MmWave massive MIMO system model.}\label{system_model}
\end{figure}

As shown in Fig.~\ref{system_model}, we consider a point-to-point massive MIMO systems working at mmWave bands, where the transmitter and the receiver are with $N_{\textrm{T}}$ and $N_{\textrm{R}}$ antennas, respectively. To reduce the hardware cost and power consumption, $N_{\textrm{T}}^{\textrm{RF}}(< N_{\textrm{T}})$ and $N_{\textrm{R}}^{\textrm{RF}}(< N_{\textrm{R}})$ RF chains are used at the transmitter and the receiver, respectively, and are connected to the large-scale antennas via phase shifters.

\subsection{Channel Model}

Due to the sparse scattering property, the Saleh-Valenzuela channel model has been used to well depict the mmWave propagation environment, where the scattering of multiple rays forms several clusters. According to \cite{O. E. Ayach}, the $N_{\textrm{R}}\times N_{\textrm{T}}$ channel matrix between the receiver and the transmitter can be represented as
\begin{eqnarray}
\label{eqn_H_tau}
\mathbf{H}=\sqrt{\frac{N_{\textrm{T}}N_{\textrm{R}}}{N_{\textrm{cl}}N_{\textrm{ray}}}}\sum_{n=1}^{N_{\textrm{cl}}}\sum_{m=1}^{N_{\textrm{ray}}} \alpha_{n,m}\mathbf{a}_{\textrm{R}}(\varphi_{n,m})\mathbf{a}_{\textrm{T}}^{H}(\phi_{n,m}),
\end{eqnarray}
where $N_{\textrm{cl}}$ and $N_{\textrm{ray}}$ denote the number of scattering clusters and the number of rays in each cluster, respectively, $\alpha_{n,m}\sim \mathcal{CN}(0, \sigma_{\alpha}^2)$ is the propagation gain of the $m$th path in the $n$th cluster with $\sigma_{\alpha}^2$ being the average power gain, $\varphi_{n,m}$ and $\phi_{n,m}\in[0,2\pi]$ are the azimuth angles of arrival and departure (AoA/AoD) at the receiver and the transmitter, respectively, of the $m$th path in the $n$th cluster.\footnote{The path gain $\alpha_{n,m}$ is the fast fading and varies in the time scale of channel coherence interval. Other parameters, $N_{\textrm{cl}}$, $N_{\textrm{ray}}$, $\varphi_{n,m}$, $\phi_{n,m}$, are slow fading and may be unchanged in a large time scale compared to $\alpha_{n,m}$. The Doppler spread determines how often these channel parameters change.} For a uniform linear array with $N$ antenna elements and an azimuth angle of $\theta$, the response vector can be expressed as
\begin{eqnarray}
\label{eqn_au}
\mathbf{a}(\theta)=\frac{1}{\sqrt{N}} \left[1,e^{-j2\pi\frac{d}{\lambda}\sin(\theta)},\ldots,e^{-j2\pi\frac{d}{\lambda}(N-1)\sin(\theta)}\right]^{T},
\end{eqnarray}
where $d$ and $\lambda$ denote the distance between the adjacent antennas and carrier wavelength, respectively.

In the above channel model, we assume the transmitted signal is with narrowband and therefore, channel matrix is independent of frequency. For wideband transmission, OFDM is used to convert a frequency-selective channel into multiple flat fading channels and the corresponding channel matrices will be different at different subcarriers. Accordingly, the design of DL-JHPF in Section III will start at the narrowband systems and is then extended to the wideband OFDM systems.

\subsection{Signal Transmission}

The transmitter sends $N_{\textrm{s}}$ parallel data streams to the receiver through the wireless channel. The bits of each data stream are first mapped to the symbol by the $M$-ary modulation. The symbol vector intended for the receiver, $\mathbf{x}\in\mathbb{C}^{N_{\textrm{s}}\times 1}$ with $\mathbb{E}\left\{\mathbf{x}\mathbf{x}^{H}\right\}=\frac{1}{N_{\textrm{s}}}\mathbf{I}_{N_{\textrm{s}}}$, is successively processed by the digital precoder, $\mathbf{F}_{\textrm{BB}}\in\mathbb{C}^{N_{\textrm{T}}^{\textrm{RF}}\times N_{\textrm{s}}}$, at the baseband and the analog precoder, $\mathbf{F}_{\textrm{RF}}\in\mathbb{C}^{N_{\textrm{T}}\times N_{\textrm{T}}^{\textrm{RF}}}$, through the phase shifters, yielding the transmitted signal
\begin{eqnarray}
\label{eqn_s}
\mathbf{s}=\sqrt{P}\mathbf{F}_{\textrm{RF}}\mathbf{F}_{\textrm{BB}}\mathbf{x},
\end{eqnarray}
where $P$ denotes the transmit power. $\mathbf{F}_{\textrm{RF}}$ represents the phase-only modulation by the phase shifters and thus has the constraint of $\left|[\mathbf{F}_{\textrm{RF}}]_{i,j}\right|=\frac{1}{\sqrt{N_{\textrm{T}}}}$, $\forall \,i, \,j$. $\mathbf{F}_{\textrm{BB}}$ is normalized as $\|\mathbf{F}_{\textrm{RF}}\mathbf{F}_{\textrm{BB}}\|_{F}^2=N_{\textrm{s}}$ to satisfy the total power constraint at the transmitter. Then the received signal at the receiver is given by
\begin{eqnarray}
\label{eqn_y}
\mathbf{y}=\sqrt{P}\mathbf{H}\mathbf{F}_{\textrm{RF}}\mathbf{F}_{\textrm{BB}}\mathbf{x}+\mathbf{n},
\end{eqnarray}
where $\mathbf{n}\in\mathbb{C}^{N_{\textrm{R}}\times 1}$ is additive white Gaussian noise (AWGN) with $\mathcal{CN}(0,1)$ elements.

The received signal $\mathbf{y}$ is then processed by the hybrid architecture at the receiver as
\setlength{\arraycolsep}{0.05em}
\begin{eqnarray}
\label{eqn_r}
\mathbf{r}&&=\mathbf{W}_{\textrm{BB}}^H\mathbf{W}_{\textrm{RF}}^H\mathbf{y} =\sqrt{P}\mathbf{W}_{\textrm{BB}}^H\mathbf{W}_{\textrm{RF}}^H\mathbf{H}\mathbf{F}_{\textrm{RF}}\mathbf{F}_{\textrm{BB}}\mathbf{x} +\mathbf{W}_{\textrm{BB}}^H\mathbf{W}_{\textrm{RF}}^H\mathbf{n},
\end{eqnarray}
where $\mathbf{W}_{\textrm{RF}}\in\mathbb{C}^{N_{\textrm{R}}\times N_{\textrm{R}}^{\textrm{RF}}}$ and $\mathbf{W}_{\textrm{BB}}\in\mathbb{C}^{N_{\textrm{R}}^{\textrm{RF}}\times N_{\textrm{s}}}$ represent the analog combiner and digital combiner, respectively. A hardware constraint is imposed on $\mathbf{W}_{\textrm{RF}}$ such that $\left|[\mathbf{W}_{\textrm{RF}}]_{i,j}\right|=\frac{1}{\sqrt{N_{\textrm{R}}}}$, $\forall \,i, \,j$ similar to $\mathbf{F}_{\textrm{RF}}$. Then the detected signal vector, $\mathbf{r}$, is demodulated to recover the original bits of $N_{\textrm{s}}$ data streams.

Since the performance of the digital communication system is ultimately determined by BER, we aim to jointly design $\mathbf{F}_{\textrm{RF}}$, $\mathbf{F}_{\textrm{BB}}$, $\mathbf{W}_{\textrm{RF}}$, and $\mathbf{W}_{\textrm{BB}}$ to minimize the BER between the original and demodulated bits, that is
\begin{eqnarray}
&&\min\limits_{\mathbf{F}_{\textrm{RF}},\mathbf{F}_{\textrm{BB}},\mathbf{W}_{\textrm{RF}},\mathbf{W}_{\textrm{BB}}}\quad P_{\textrm{e}}\left(\mathbf{F}_{\textrm{RF}},\mathbf{F}_{\textrm{BB}},\mathbf{W}_{\textrm{RF}},\mathbf{W}_{\textrm{BB}}\right),\label{eqn_optim_problem}\\
&&\quad\quad\:\:\textrm{s.t.}\qquad\qquad\!\! \left|[\mathbf{F}_{\textrm{RF}}]_{i,j}\right|=\frac{1}{\sqrt{N_{\textrm{T}}}}, \forall \,i, \,j,\label{const1}\\
&&\qquad\qquad\qquad\quad\;\; \left|[\mathbf{W}_{\textrm{RF}}]_{i,j}\right|=\frac{1}{\sqrt{N_{\textrm{R}}}}, \forall \,i, \,j,\label{const2}\\
&&\qquad\qquad\qquad\quad\;\; \|\mathbf{F}_{\textrm{RF}}\mathbf{F}_{\textrm{BB}}\|_{F}^2=N_{\textrm{s}}\label{const3}.
\end{eqnarray}
The BER in (\ref{eqn_optim_problem}) is a complicated nonlinear function of $\mathbf{F}_{\textrm{RF}}$, $\mathbf{F}_{\textrm{BB}}$, $\mathbf{W}_{\textrm{RF}}$, and $\mathbf{W}_{\textrm{BB}}$ without closed-form expression and the constraints in (\ref{const1}) and (\ref{const2}) are non-convex, which make this optimization problem intractable to be solved by the traditional approaches. DL is a potential solution by using the BP algorithm and thus we develop DL-JHPF to address this problem.

\begin{figure*}[t]
\centering
\includegraphics[width=6.6in]{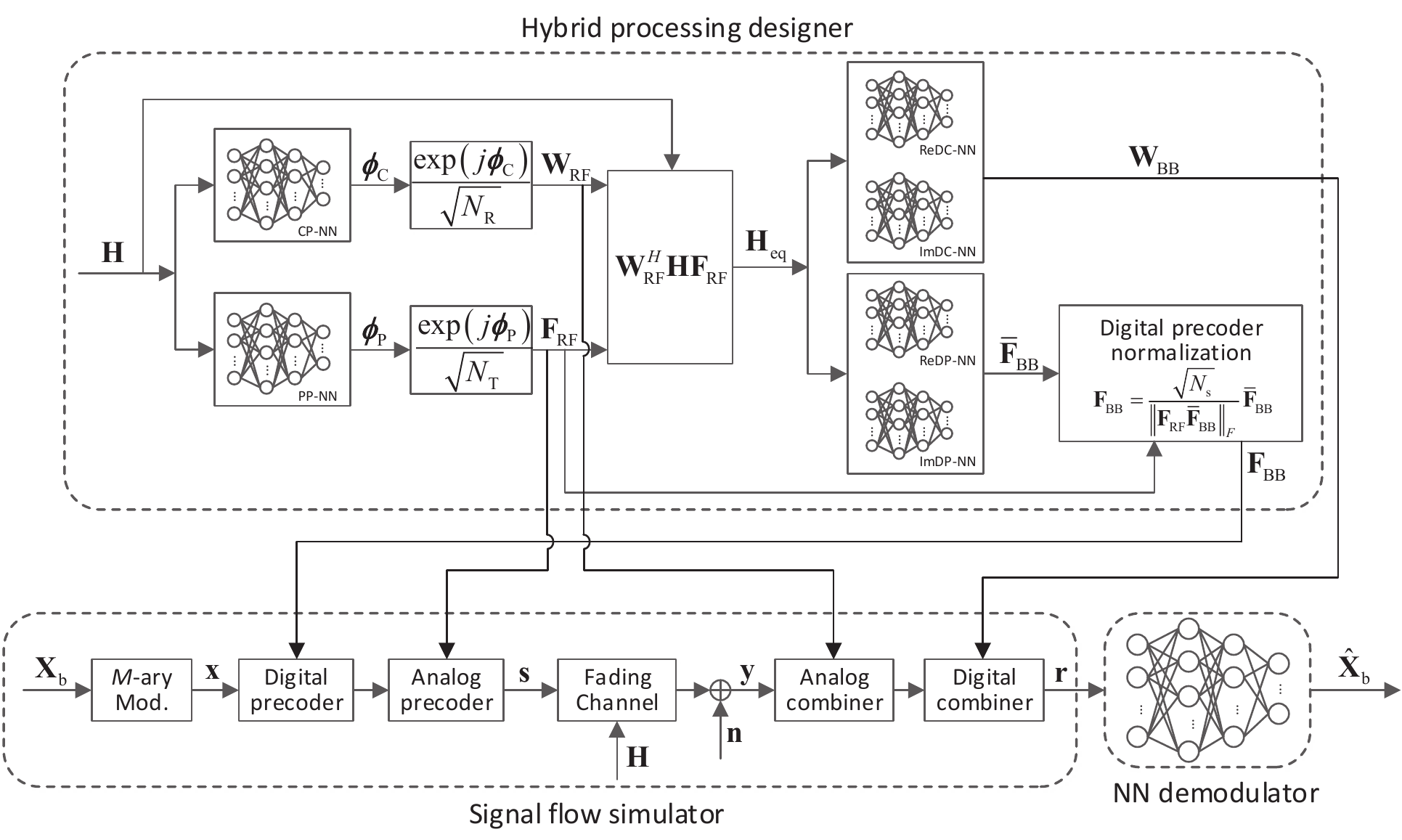}
\caption{Proposed DL-JHPF.}\label{DL_framework}
\end{figure*}

\section{Proposed DL-JHPF}

In this section, we first briefly review the existing work on the DNN based end-to-end communications. Then we propose DL-JHPF, where the framework is first described, followed by the details of training, deployment, and testing along with the corresponding complexity analysis. Finally, we extend the framework to OFDM systems over wideband mmWave channels.

\subsection{DNN based End-to-End Communications}

Prior works have shown that DNN based end-to-end optimization is an efficient tool to minimize BER. The BP algorithm makes the DNN based end-to-end communications over the air possible so long as the optimized performance metric is differentiable \cite{T. O'hea}, \cite{S. Dorner}, \cite{H. Ye_b}. For the DNN based end-to-end communication system, the modules at the transmitter and the receiver are replaced by two DNNs, respectively. Specifically, the DNN at the transmitter encodes the original symbols into the transmitted signal and the one at the receiver recovers the original symbols from the output of the wireless channel. In the training stage, the error between the original and recovered symbols is computed and the weights of the two DNNs are adjusted iteratively based on the error gradient propagated from the output layer of the DNN at the receiver to optimize the recovery accuracy.

In this paper, we focus on the DL based joint analog and digital processing design for the transceiver in mmWave massive MIMO systems. Then, the existing DNN based end-to-end communication is not suitable for this task since it integrates the modules of the transceiver into two DNNs and thus cannot meet the hardware and power constraints in practical implementation. To address this challenge, we design DL-JHPF in the following.

\subsection{Framework Description}


As shown in Fig.~\ref{DL_framework}, the proposed DL-JHPF consists of three parts: hybrid processing designer, signal flow simulator, and NN demodulator, which are elaborated as follows.

\emph{Hybrid processing designer:} It plays the role of outputting the hybrid processing matrices for the transceiver by using NNs based on the channel matrix. It includes six fully-connected NNs and is used to generate the analog and digital processing matrices for the transmitter and the receiver based on the channel matrix, $\mathbf{H}$. Specifically, $\mathbf{H}\in\mathbb{C}^{N_{\textrm{R}}\times N_{\textrm{T}}}$ is first converted to a $2N_{\textrm{T}}N_{\textrm{R}}\times 1$ real-valued vector.\footnote{In Fig.~\ref{DL_framework}, only the main process of the framework is shown while the matrix and vector reshaping process is omitted.} Then it is input into two NNs, called precoder phase NN (PP-NN) and combiner phase NN (CP-NN), to generate the corresponding phases, $\boldsymbol{\phi}_{\textrm{P}}\in\mathbb{R}^{N_{\textrm{T}}N_{\textrm{T}}^{\textrm{RF}}\times 1}$ and $\boldsymbol{\phi}_{\textrm{C}}\in\mathbb{R}^{N_{\textrm{R}}N_{\textrm{R}}^{\textrm{RF}}\times 1}$, respectively, for phase shifters. With $\boldsymbol{\phi}_{\textrm{P}}$ and $\boldsymbol{\phi}_{\textrm{C}}$, two complex-valued vectors with constant amplitude elements are generated as
\begin{eqnarray}
\label{eqn_fRF}
\bar{\mathbf{f}}_{\textrm{RF}}=\frac{1}{\sqrt{N_{\textrm{T}}}}e^{j\boldsymbol{\phi}_{\textrm{P}}},
\end{eqnarray}
\begin{eqnarray}
\label{eqn_wRF}
\bar{\mathbf{w}}_{\textrm{RF}}=\frac{1}{\sqrt{N_{\textrm{R}}}}e^{j\boldsymbol{\phi}_{\textrm{C}}},
\end{eqnarray}
based on which, $\mathbf{F}_{\textrm{RF}}$ and $\mathbf{W}_{\textrm{RF}}$ are given by
\begin{eqnarray}
\label{eqn_FRF}
\mathbf{F}_{\textrm{RF}}=\mathcal{T}_{\textrm{v}\rightarrow \textrm{m}}(\bar{\mathbf{f}}_{\textrm{RF}}),
\end{eqnarray}
\begin{eqnarray}
\label{eqn_WRF}
\mathbf{W}_{\textrm{RF}}=\mathcal{T}_{\textrm{v}\rightarrow \textrm{m}}(\bar{\mathbf{w}}_{\textrm{RF}}),
\end{eqnarray}
where $\mathcal{T}_{\textrm{v}\rightarrow \textrm{m}}(\cdot)$ denotes the operation reshaping a vector to a matrix.
Then, $\mathbf{F}_{\textrm{RF}}$ and $\mathbf{W}_{\textrm{RF}}$ along with $\mathbf{H}$ are used to generate a low-dimensional equivalent channel, i.e.,
\begin{eqnarray}
\label{eqn_Heq}
\mathbf{H}_{\textrm{eq}}=\mathbf{W}_{\textrm{RF}}^H\mathbf{H}\mathbf{F}_{\textrm{RF}}.
\end{eqnarray}
$\mathbf{H}_{\textrm{eq}}\in\mathbb{C}^{N_{\textrm{R}}^{\textrm{RF}}\times N_{\textrm{T}}^{\textrm{RF}}}$ is converted to a $2N_{\textrm{T}}^{\textrm{RF}}N_{\textrm{R}}^{\textrm{RF}}\times 1$ real-valued vector before it is input into four parallel NNs. The first two NNs, corresponding to the real part digital combiner NN (ReDC-NN) and the imaginary part digital combiner NN (ImDC-NN), output two $N_{\textrm{s}}N_{\textrm{R}}^{\textrm{RF}}\times1$ vectors, $\bar{\mathbf{w}}_{\textrm{BB,re}}$, $\bar{\mathbf{w}}_{\textrm{BB,im}}$, respectively. Then $\mathbf{W}_{\textrm{BB}}$ can be obtained as
\begin{eqnarray}
\label{eqn_WBB}
\mathbf{W}_{\textrm{BB}}=\mathcal{T}_{\textrm{v}\rightarrow \textrm{m}}(\bar{\mathbf{w}}_{\textrm{BB,re}}+j\bar{\mathbf{w}}_{\textrm{BB,im}}).
\end{eqnarray}
Another two NNs, corresponding to the real part digital precoder NN (ReDP-NN) and the imaginary part digital precoder NN (ImDP-NN), output two $N_{\textrm{s}}N_{\textrm{T}}^{\textrm{RF}}\times1$ vectors, $\bar{\mathbf{f}}_{\textrm{BB,re}}$, $\bar{\mathbf{f}}_{\textrm{BB,im}}$, respectively. Then the unnormalized digital precoder $\bar{\mathbf{F}}_{\textrm{BB}}$ is given by
\begin{eqnarray}
\label{eqn_unnorm_FBB}
\bar{\mathbf{F}}_{\textrm{BB}}=\mathcal{T}_{\textrm{v}\rightarrow \textrm{m}}(\bar{\mathbf{f}}_{\textrm{BB,re}}+j\bar{\mathbf{f}}_{\textrm{BB,im}}).
\end{eqnarray}
The following normalization utilizes $\bar{\mathbf{F}}_{\textrm{BB}}$ and $\mathbf{F}_{\textrm{RF}}$ in (\ref{eqn_FRF}) to output the final digital precoder as
\begin{eqnarray}
\label{eqn_DP_norm}
\mathbf{F}_{\textrm{BB}}=\frac{\sqrt{N_{\textrm{s}}}}{\|\mathbf{F}_{\textrm{RF}}\bar{\mathbf{F}}_{\textrm{BB}}\|_{F}}\bar{\mathbf{F}}_{\textrm{BB}}.
\end{eqnarray}

\emph{Signal flow simulator:} In the training stage, it simulates the process from the original bits, $\mathbf{X}_{\textrm{b}}$, to the detected signal, $\mathbf{r}$, over the channel, $\mathbf{H}$, with AWGN, $\mathbf{n}$, where $\mathbf{X}_{\textrm{b}}$ with the size of $N_{\textrm{s}}\times\log_2 M$, $\mathbf{H}$, and $\mathbf{n}$ are generated in the simulation environment. It bridges the back propagation of the error gradient from NN demodulator to hybrid processing designer as we will elaborate in Section III.C. In the deployment and testing stage, the signal flow simulator is replaced by the actual transceiver and the actual wireless fading channel. In these two stages, the analog and digital processing matrices at the transceiver are provided by the hybrid processing designer based on the simulated or actual $\mathbf{H}$.

\emph{NN demodulator:} It is a fully-connected NN, which receives the detected signal, $\mathbf{r}$, from the signal flow simulator (in the training stage) or the actual receiver (in the testing stage) and outputs recovered bits $\hat{\mathbf{x}}_{\textrm{b}}\in \mathbb{R}^{N_{\textrm{s}}\log_2 M\times1}$ with each element lies in the interval $[0,1]$. $\hat{\mathbf{x}}_{\textrm{b}}$ is then reshaped to $\hat{\mathbf{X}}_{\textrm{b}}$ with the same size as $\mathbf{X}_{\textrm{b}}$.

\textbf{Remark 1.} \emph{The learning of hybrid processing matrices, $\mathbf{F}_{\textrm{RF}}$, $\mathbf{W}_{\textrm{RF}}$, $\mathbf{F}_{\textrm{BB}}$, and $\mathbf{W}_{\textrm{BB}}$, in DL-JHPF is embedded into the signal transmission and demodulation process instead of approximating pre-designed label matrices. All NNs are optimized jointly sharing the mapping principle from $\mathbf{X}_{\textrm{b}}$ at the transmitter to $\mathbf{X}_{\textrm{b}}$ at the receiver that resembles an
autoencoder. By minimizing the error between $\mathbf{X}_{\textrm{b}}$ and $\hat{\mathbf{X}}_{\textrm{b}}$, each NN in hybrid processing designer can learn to output the appropriate vectors with specific meaning implicitly, i.e., phases of phase shifters and real and imaginary parts of the digital precoder and combiner. By doing this, DL-JHPF will have the potential to break through the performance of the existing schemes.}

\subsection{Framework Training}

The goal of offline training is to determine the weights of the NNs in hybrid processing designer and NN demodulator based on the training samples with the input tuple $\langle\mathbf{H}, \mathbf{X}_{\textrm{b}}, \mathbf{n}\rangle$ and the label $\mathbf{X}_{\textrm{b}}$, where $\mathbf{H}$ is generated by certain channel model and $\mathbf{n}$ is generated according to the $\mathcal{CN}(0,1)$ distribution. By minimizing the end-to-end error between the original bits, $\mathbf{X}_{\textrm{b}}$, and the recovered bits, $\hat{\mathbf{X}}_{\textrm{b}}$, the weights of each NN in DL-JHPF are adjusted iteratively and the training procedure is elaborated as follows.

\begin{figure}[t]
\centering
\includegraphics[width=3.0in]{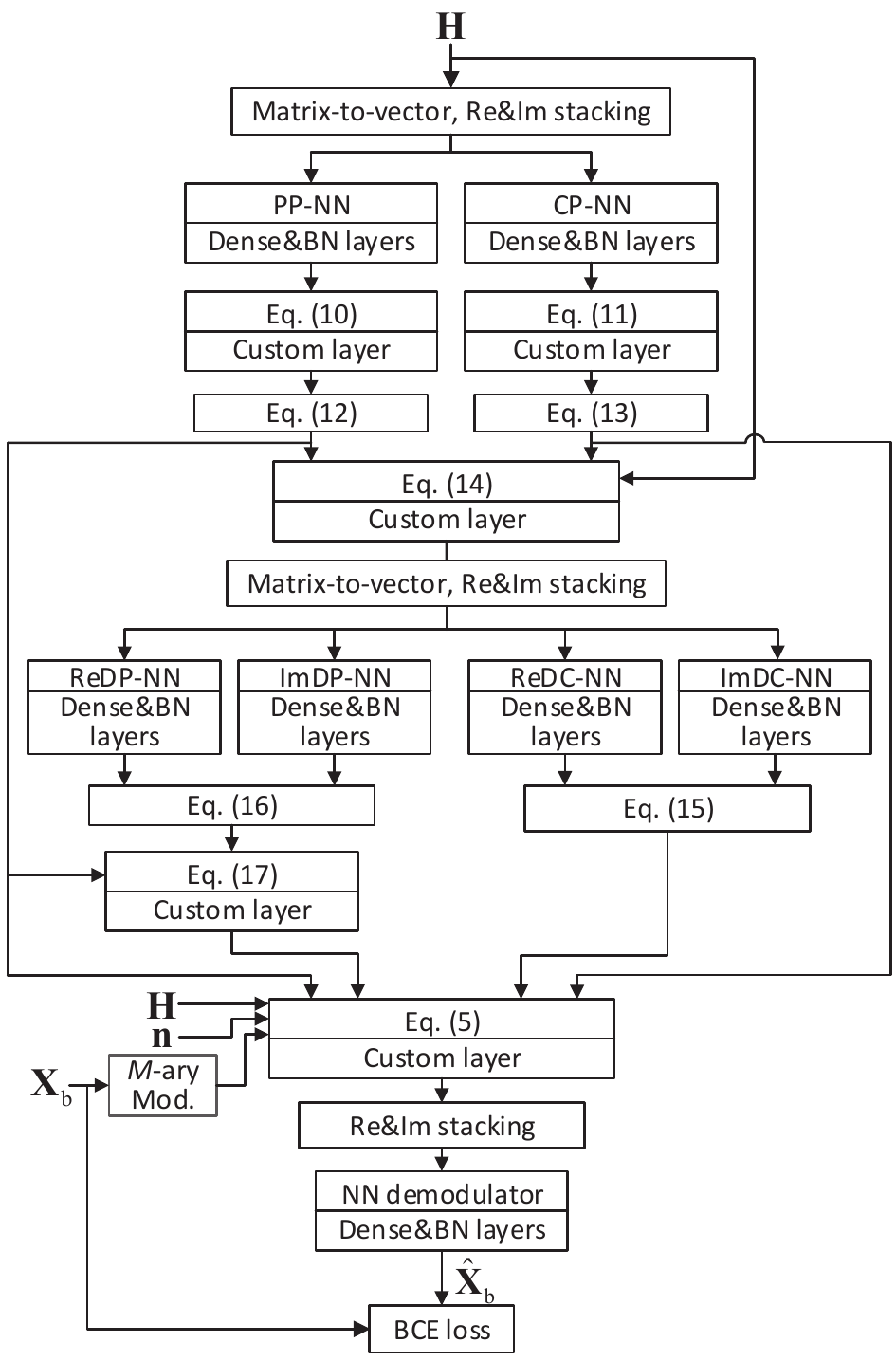}
\caption{Training model for proposed DL-JHPF.}\label{visual_model}
\end{figure}

The proposed DL-JHPF is actually an integrated DNN consisting of neuron layers and custom layers. The training model in Fig.~\ref{visual_model} demonstrates the detailed training process of the framework. For each training sample, $\mathbf{H}$ is converted into a real-valued vector by matrix-to-vector reshaping and real and imaginary parts stacking, which is input into PP-NN and CP-NN consisting of dense and batch normalization (BN) layers to generate the corresponding phases, $\boldsymbol{\phi}_{\textrm{P}}$ and $\boldsymbol{\phi}_{\textrm{C}}$, respectively. Then (\ref{eqn_fRF}) and (\ref{eqn_wRF}) are executed by the same custom layer. Afterwards, the output vectors are reshaped according to (\ref{eqn_FRF}) and (\ref{eqn_WRF}) to generate $\mathbf{F}_{\textrm{RF}}$ and $\mathbf{W}_{\textrm{RF}}$, respectively. Next, (\ref{eqn_Heq}) is executed by a custom layer to generate $\mathbf{H}_{\textrm{eq}}$, followed by matrix-to-vector reshaping and real and imaginary parts stacking. This  vector is input into four NNs consisting of dense and BN layers, i.e., ReDC-NN, ImDC-NN, ReDP-NN, and ImDP-NN, respectively. The output vectors of the former two NNs are used to generate $\mathbf{W}_{\textrm{BB}}$ through real and imaginary parts combining and vector-to-matrix reshaping as (\ref{eqn_WBB}). Using the same operation, the output vectors of the latter two NNs are used to generate $\bar{\mathbf{F}}_{\textrm{BB}}$ as (\ref{eqn_unnorm_FBB}). After obtaining $\bar{\mathbf{F}}_{\textrm{BB}}$, a custom layer is added to perform the normalization in (\ref{eqn_DP_norm}) to generate $\mathbf{F}_{\textrm{BB}}$. Then (\ref{eqn_r}) is executed through a custom layer by using the input tuple $\langle\mathbf{H}, \mathbf{X}_{\textrm{b}}, \mathbf{n}\rangle$ and the generated $\mathbf{F}_{\textrm{RF}}$, $\mathbf{F}_{\textrm{BB}}$, $\mathbf{W}_{\textrm{RF}}$, and $\mathbf{W}_{\textrm{BB}}$ to yield the detected signal, $\mathbf{r}$. After real and imaginary stacking, $\mathbf{r}$ is converted to a real-valued vector and input into the NN demodulator consisting of dense and BN layers to output the recovered bits, $\hat{\mathbf{x}}_{\textrm{b}}$, which is then reshaped to $\hat{\mathbf{X}}_{\textrm{b}}$. The binary cross-entropy (BCE) loss between $\mathbf{X}_{\textrm{b}}$ and $\hat{\mathbf{X}}_{\textrm{b}}$ is calculated as
\begin{eqnarray}
\label{eqn_loss}
\mathcal{L}&&=-\frac{1}{N_{\textrm{tr}}}\sum_{n=1}^{N_{\textrm{tr}}}\sum_{i=1}^{N_{\textrm{s}}}\sum_{j=1}^{\log_2 M}\biggl( [\mathbf{X}_{\textrm{b}}^{n}]_{i,j}\ln([\hat{\mathbf{X}}_{\textrm{b}}^{n}(\boldsymbol\Theta)]_{i,j})\nonumber\\
&&\quad+(1-[\mathbf{X}_{\textrm{b}}^{n}]_{i,j})\ln(1-[\hat{\mathbf{X}}_{\textrm{b}}^{n}(\boldsymbol\Theta)]_{i,j})\biggr),
\end{eqnarray}
where $N_{\textrm{tr}}$ denotes the number of training samples, superscript $n$ is added to indicate the index of the training sample, and $\hat{\mathbf{X}}_{\textrm{b}}^{n}$ is expressed as the function of the parameter set of all NNs in DL-JHPF, i.e., $\boldsymbol\Theta$.

Recall the optimization problem in (\ref{eqn_optim_problem}), the BER over the training set can be written as
\begin{eqnarray}
\label{eqn_BER}
&&P_{\textrm{e,tr}}\left(\mathbf{F}_{\textrm{RF}},\mathbf{F}_{\textrm{BB}},\mathbf{W}_{\textrm{RF}},\mathbf{W}_{\textrm{BB}}\right) =P_{\textrm{e,tr}}\left(\boldsymbol\Theta\right) \nonumber\\
&&\quad=\frac{\sum_{n=1}^{N_{\textrm{tr}}}\sum_{i=1}^{N_{\textrm{s}}}\sum_{j=1}^{\log_2 M}|[\mathbf{X}_{\textrm{b}}^{n}]_{i,j}-[\hat{\mathbf{X}}_{\textrm{b,bin}}^{n}(\boldsymbol\Theta)]_{i,j}|}{N_{\textrm{tr}}N_{\textrm{s}}\log_2 M},\;\;\;\;
\end{eqnarray}
where $\hat{\mathbf{X}}_{\textrm{b,bin}}^{n}(\boldsymbol\Theta)$ is the binary demodulated bit matrix with $[\hat{\mathbf{X}}_{\textrm{b,1hot}}^{n}(\boldsymbol\Theta)]_{i,j}=0$ for $[\hat{\mathbf{X}}_{\textrm{b}}^{n}(\boldsymbol\Theta)]_{i,j}<0.5$ and $1$ otherwise. With $\mathbf{X}_{\textrm{b}}^{n}$ fixed, minimizing $\mathcal{L}$ in (\ref{eqn_loss}) with respect to $\hat{\mathbf{X}}_{\textrm{b}}^{n}(\boldsymbol\Theta)$ yields $\hat{\mathbf{X}}_{\textrm{b}}^{n}(\boldsymbol\Theta)=\mathbf{X}_{\textrm{b}}^{n}$, which also minimizes $P_{\textrm{e,tr}}\left(\boldsymbol\Theta\right)$ in (\ref{eqn_BER}). Therefore, DL-JHPF can directly minimize the BER over the training set by minimizing the BCE loss and the feasibility is guaranteed by the following theorem.

\textbf{Theorem 1.} \emph{The proposed DL-JHPF is trainable and can minimize the BCE loss through BP algorithm.}
\begin{IEEEproof}
Considering the mini-batch training, the BCE loss over a batch is written as
\begin{eqnarray}
\label{eqn_loss_batch}
\mathcal{L}_{\textrm{bat}}&&=-\frac{1}{N_{\textrm{bat}}}\sum_{n=1}^{N_{\textrm{bat}}}\sum_{i=1}^{N_{\textrm{s}}}\sum_{j=1}^{\log_2 M}\biggl( [\mathbf{X}_{\textrm{b}}^{n}]_{i,j}\ln([\hat{\mathbf{X}}_{\textrm{b}}^{n}(\boldsymbol\Theta)]_{i,j})\nonumber\\
&&\quad+(1-[\mathbf{X}_{\textrm{b}}^{n}]_{i,j})\ln(1-[\hat{\mathbf{X}}_{\textrm{b}}^{n}(\boldsymbol\Theta)]_{i,j})\biggr),
\end{eqnarray}
where $N_{\textrm{bat}}$ denotes the batch size. Then $\boldsymbol\Theta$ will be updated $\lceil\frac{N_{\textrm{tr}}}{N_{\textrm{bat}}}\rceil$ times in each epoch.

To prove Theorem 1, we need to show that $\mathcal{L}_{\textrm{bat}}$ is differentiable with respect to each parameter in $\boldsymbol\Theta$. According to \cite{Y. LeCun}, the outputs are differentiable with respect to the corresponding weights and inputs for each NN in DL-JHPF. Since DL-JHPF can be viewed as an integrated DNN consisting of neuron layers and custom layers, the proof can be further simplified to prove that $\mathcal{L}_{\textrm{bat}}$ is differentiable with respect to the outputs of each NN due to chain rule. In the following, we prove the differentiability of $\mathcal{L}_{\textrm{bat}}$ with respect to the outputs of each NN by incorporating the custom layers.

\emph{NN demodulator:} From (\ref{eqn_loss_batch}), $\mathcal{L}_{\textrm{bat}}$ is differentiable with respect to $[\hat{\mathbf{X}}_{\textrm{b}}^{n}(\boldsymbol\Theta)]_{i,j}, \forall i, j$.

\emph{Re/ImDC-NN:} As mentioned in Section III.B, $\bar{\mathbf{w}}_{\textrm{BB,re}}$ and $\bar{\mathbf{w}}_{\textrm{BB,im}}$ are the outputs of ReDC-NN and ImDC-NN, respectively. Without loss of generality, we will prove that $\mathcal{L}_{\textrm{bat}}$ is differentiable with respect to $[\bar{\mathbf{w}}_{\textrm{BB,re}}]_{1}$ and $[\bar{\mathbf{w}}_{\textrm{BB,im}}]_{1}$. According to (\ref{eqn_r}) and (\ref{eqn_WBB}), $[\mathbf{r}]_{1}$ is the function of $[\bar{\mathbf{w}}_{\textrm{BB,re}}]_{1}$ and $[\bar{\mathbf{w}}_{\textrm{BB,im}}]_{1}$, that is
\begin{eqnarray}
\label{eqn_r1}
\!\!\!\!\![\mathbf{r}]_{1}&&=([\bar{\mathbf{w}}_{\textrm{BB,re}}]_{1}-j[\bar{\mathbf{w}}_{\textrm{BB,im}}]_{1})[\mathbf{z}]_{1}+C_1\nonumber\\
&&=([\bar{\mathbf{w}}_{\textrm{BB,re}}]_{1}-j[\bar{\mathbf{w}}_{\textrm{BB,im}}]_{1})([\mathbf{z}]_{1,\textrm{re}}+j[\mathbf{z}]_{1,\textrm{im}})+C_1\nonumber\\
&&=([\bar{\mathbf{w}}_{\textrm{BB,re}}]_{1}[\mathbf{z}]_{1,\textrm{re}}+[\bar{\mathbf{w}}_{\textrm{BB,im}}]_{1}[\mathbf{z}]_{1,\textrm{im}}+C_{1,\textrm{re}})\nonumber\\
&&\quad+j([\bar{\mathbf{w}}_{\textrm{BB,re}}]_{1}[\mathbf{z}]_{1,\textrm{im}}-[\bar{\mathbf{w}}_{\textrm{BB,im}}]_{1}[\mathbf{z}]_{1,\textrm{re}}+C_{1,\textrm{im}})\nonumber\\
&&=[\mathbf{r}]_{1,\textrm{re}}+j[\mathbf{r}]_{1,\textrm{im}},
\end{eqnarray}
where $\mathbf{z}=\mathbf{W}_{\textrm{RF}}^H\mathbf{y}$ and $C_1$ denotes the component of $[\mathbf{r}]_{1}$ independent of $[\bar{\mathbf{w}}_{\textrm{BB,re}}]_{1}$ and $[\bar{\mathbf{w}}_{\textrm{BB,im}}]_{1}$ with the subscripts `re' and `im' indicating the real and imaginary parts, respectively. Since $[\mathbf{r}]_{1,\textrm{re}}$ and $[\mathbf{r}]_{1,\textrm{im}}$ are a part of inputs of NN demodulator, $\mathcal{L}_{\textrm{bat}}$ is differentiable with respect to $[\mathbf{r}]_{1,\textrm{re}}$ and $[\mathbf{r}]_{1,\textrm{im}}$. Then we have
\begin{eqnarray}
\label{eqn_derL_wre_1}
\frac{\partial\mathcal{L}_{\textrm{bat}}}{\partial[\bar{\mathbf{w}}_{\textrm{BB,re}}]_{1}} &&=\frac{\partial\mathcal{L}_{\textrm{bat}}}{\partial[\mathbf{r}]_{1,\textrm{re}}} \frac{\partial[\mathbf{r}]_{1,\textrm{re}}}{\partial[\bar{\mathbf{w}}_{\textrm{BB,re}}]_{1}} +\frac{\partial\mathcal{L}_{\textrm{bat}}}{\partial[\mathbf{r}]_{1,\textrm{im}}} \frac{\partial[\mathbf{r}]_{1,\textrm{im}}}{\partial[\bar{\mathbf{w}}_{\textrm{BB,re}}]_{1}}\nonumber\\
&&=[\mathbf{z}]_{1,\textrm{re}}\frac{\partial\mathcal{L}_{\textrm{bat}}}{\partial[\mathbf{r}]_{1,\textrm{re}}} +[\mathbf{z}]_{1,\textrm{im}}\frac{\partial\mathcal{L}_{\textrm{bat}}}{\partial[\mathbf{r}]_{1,\textrm{im}}},
\end{eqnarray}
\begin{eqnarray}
\label{eqn_derL_wim_1}
\frac{\partial\mathcal{L}_{\textrm{bat}}}{\partial[\bar{\mathbf{w}}_{\textrm{BB,im}}]_{1}} &&=\frac{\partial\mathcal{L}_{\textrm{bat}}}{\partial[\mathbf{r}]_{1,\textrm{re}}} \frac{\partial[\mathbf{r}]_{1,\textrm{re}}}{\partial[\bar{\mathbf{w}}_{\textrm{BB,im}}]_{1}} +\frac{\partial\mathcal{L}_{\textrm{bat}}}{\partial[\mathbf{r}]_{1,\textrm{im}}} \frac{\partial[\mathbf{r}]_{1,\textrm{im}}}{\partial[\bar{\mathbf{w}}_{\textrm{BB,im}}]_{1}}\nonumber\\
&&=[\mathbf{z}]_{1,\textrm{im}}\frac{\partial\mathcal{L}_{\textrm{bat}}}{\partial[\mathbf{r}]_{1,\textrm{re}}} -[\mathbf{z}]_{1,\textrm{re}}\frac{\partial\mathcal{L}_{\textrm{bat}}}{\partial[\mathbf{r}]_{1,\textrm{im}}}.
\end{eqnarray}

\emph{Re/ImDP-NN:} Since $\bar{\mathbf{f}}_{\textrm{BB,re}}$ and $\bar{\mathbf{f}}_{\textrm{BB,im}}$ are the outputs of ReDP-NN and ImDP-NN, respectively, we also aim to prove that $\mathcal{L}_{\textrm{bat}}$ is differentiable with respect to $[\bar{\mathbf{f}}_{\textrm{BB,re}}]_{1}$ and $[\bar{\mathbf{f}}_{\textrm{BB,im}}]_{1}$. Considering the normalization in (\ref{eqn_DP_norm}), we first calculate the derivatives of $\mathcal{L}_{\textrm{bat}}$ with respect to the real and imaginary parts of $[\mathbf{F}_{\textrm{BB}}]_{1,1}$, i.e., $\frac{\partial\mathcal{L}_{\textrm{bat}}}{\partial[\mathbf{F}_{\textrm{BB}}]_{1,1,\textrm{re}}}$ and $\frac{\partial\mathcal{L}_{\textrm{bat}}}{\partial[\mathbf{F}_{\textrm{BB}}]_{1,1,\textrm{im}}}$, which can be obtained similarly to (\ref{eqn_derL_wre_1}) and (\ref{eqn_derL_wim_1}). According to (\ref{eqn_DP_norm}), we have
\begin{eqnarray}
\label{eqn_FBB11re}
[\mathbf{F}_{\textrm{BB}}]_{1,1,\textrm{re}}=\frac{[\bar{\mathbf{f}}_{\textrm{BB,re}}]_{1}}{f([\bar{\mathbf{f}}_{\textrm{BB,re}}]_{1}, [\bar{\mathbf{f}}_{\textrm{BB,im}}]_{1})},
\end{eqnarray}
\begin{eqnarray}
\label{eqn_FBB11im}
[\mathbf{F}_{\textrm{BB}}]_{1,1,\textrm{im}}=\frac{[\bar{\mathbf{f}}_{\textrm{BB,im}}]_{1}}{f([\bar{\mathbf{f}}_{\textrm{BB,re}}]_{1}, [\bar{\mathbf{f}}_{\textrm{BB,im}}]_{1})},
\end{eqnarray}
where $f([\bar{\mathbf{f}}_{\textrm{BB,re}}]_{1}, [\bar{\mathbf{f}}_{\textrm{BB,im}}]_{1})\!=\![([\mathbf{F}_{\textrm{RF}}]_{1,1,\textrm{re}}[\bar{\mathbf{f}}_{\textrm{BB,re}}]_{1} \!-\![\mathbf{F}_{\textrm{RF}}]_{1,1,\textrm{im}}[\bar{\mathbf{f}}_{\textrm{BB,im}}]_{1}+C_{2,\textrm{re}})^2+([\mathbf{F}_{\textrm{RF}}]_{1,1,\textrm{re}}[\bar{\mathbf{f}}_{\textrm{BB,im}}]_{1} \!+\![\mathbf{F}_{\textrm{RF}}]_{1,1,\textrm{im}}[\bar{\mathbf{f}}_{\textrm{BB,re}}]_{1}+C_{2,\textrm{im}})^2+C_3]^{\frac{1}{2}}$ with $C_2$ and $C_3$ independent of $[\bar{\mathbf{f}}_{\textrm{BB,re}}]_{1}$ and $[\bar{\mathbf{f}}_{\textrm{BB,im}}]_{1}$. Then we can find that $[\mathbf{F}_{\textrm{BB}}]_{1,1,\textrm{re}}$ and $[\mathbf{F}_{\textrm{BB}}]_{1,1,\textrm{im}}$ are differentiable with respect to $[\bar{\mathbf{f}}_{\textrm{BB,re}}]_{1}$ and $[\bar{\mathbf{f}}_{\textrm{BB,im}}]_{1}$, which leads to
\begin{eqnarray}
\label{eqn_derL_fBBre_1}
\frac{\partial\mathcal{L}_{\textrm{bat}}}{\partial[\bar{\mathbf{f}}_{\textrm{BB,re}}]_{1}} &&\!=\!\frac{\partial\mathcal{L}_{\textrm{bat}}}{[\mathbf{F}_{\textrm{BB}}]_{1,1,\textrm{re}}} \frac{\partial[\mathbf{F}_{\textrm{BB}}]_{1,1,\textrm{re}}}{\partial[\bar{\mathbf{f}}_{\textrm{BB,re}}]_{1}} \!+\!\frac{\partial\mathcal{L}_{\textrm{bat}}}{\partial[\mathbf{F}_{\textrm{BB}}]_{1,1,\textrm{im}}} \frac{\partial[\mathbf{F}_{\textrm{BB}}]_{1,1,\textrm{im}}}{\partial[\bar{\mathbf{f}}_{\textrm{BB,re}}]_{1}},\nonumber\\
&&
\end{eqnarray}
\begin{eqnarray}
\label{eqn_derL_fBBim_1}
\frac{\partial\mathcal{L}_{\textrm{bat}}}{\partial[\bar{\mathbf{f}}_{\textrm{BB,im}}]_{1}} &&\!=\!\frac{\partial\mathcal{L}_{\textrm{bat}}}{[\mathbf{F}_{\textrm{BB}}]_{1,1,\textrm{re}}} \frac{\partial[\mathbf{F}_{\textrm{BB}}]_{1,1,\textrm{re}}}{\partial[\bar{\mathbf{f}}_{\textrm{BB,im}}]_{1}} \!+\!\frac{\partial\mathcal{L}_{\textrm{bat}}}{\partial[\mathbf{F}_{\textrm{BB}}]_{1,1,\textrm{im}}} \frac{\partial[\mathbf{F}_{\textrm{BB}}]_{1,1,\textrm{im}}}{\partial[\bar{\mathbf{f}}_{\textrm{BB,im}}]_{1}}.\nonumber\\
&&
\end{eqnarray}

\emph{PP-NN:} We still aim to prove that $\mathcal{L}_{\textrm{bat}}$ is differentiable with respect to $[\boldsymbol{\phi}_{\textrm{P}}]_{1}$ that is one of the output of PP-NN and generates $[\bar{\mathbf{f}}_{\textrm{RF}}]_{1,\textrm{re}}$ and $[\bar{\mathbf{f}}_{\textrm{RF}}]_{1,\textrm{im}}$ as
\begin{eqnarray}
\label{eqn_fRF_reim}
\frac{1}{\sqrt{N_{\textrm{T}}}}e^{j[\boldsymbol{\phi}_{\textrm{P}}]_{1}}&&=\frac{1}{\sqrt{N_{\textrm{T}}}}(\cos[\boldsymbol{\phi}_{\textrm{P}}]_{1} +j\sin[\boldsymbol{\phi}_{\textrm{P}}]_{1})\nonumber\\
&&=[\bar{\mathbf{f}}_{\textrm{RF}}]_{1,\textrm{re}}+j[\bar{\mathbf{f}}_{\textrm{RF}}]_{1,\textrm{im}}.
\end{eqnarray}
From (\ref{eqn_r}) and (\ref{eqn_Heq}), $[\bar{\mathbf{f}}_{\textrm{RF}}]_{1,\textrm{re}}$ and $[\bar{\mathbf{f}}_{\textrm{RF}}]_{1,\textrm{im}}$ influence the values of $[\mathbf{r}]_{i}, i=1,\ldots,N_{\textrm{s}}$ and $[\mathbf{H}_{\textrm{eq}}]_{j,1}, j=1,\ldots,N_{\textrm{R}}^{\textrm{RF}}$. According to the previous proof, $\mathcal{L}_{\textrm{bat}}$ is differentiable with respect to the real and imaginary parts of each element in $\mathbf{r}$ and $\mathbf{H}_{\textrm{eq}}$. $[\mathbf{r}]_{i,\textrm{re}}$, $[\mathbf{r}]_{i,\textrm{im}}$, $[\mathbf{H}_{\textrm{eq}}]_{j,1,\textrm{re}}$, and $[\mathbf{H}_{\textrm{eq}}]_{j,1,\textrm{im}}$, $i=1,\ldots,N_{\textrm{s}}, j=1,\ldots,N_{\textrm{R}}^{\textrm{RF}}$ are also differentiable with respect to $[\bar{\mathbf{f}}_{\textrm{RF}}]_{1,\textrm{re}}$ and $[\bar{\mathbf{f}}_{\textrm{RF}}]_{1,\textrm{im}}$. Resorting to chain rule, we have
\begin{eqnarray}
\label{eqn_derL_fRF_1re}
&&\frac{\partial\mathcal{L}_{\textrm{bat}}}{\partial[\bar{\mathbf{f}}_{\textrm{RF}}]_{1,\textrm{re}}} =\sum_{i=1}^{N_{\textrm{s}}}\left(\frac{\partial\mathcal{L}_{\textrm{bat}}}{\partial[\mathbf{r}]_{i,\textrm{re}}} \frac{\partial[\mathbf{r}]_{i,\textrm{re}}}{\partial[\bar{\mathbf{f}}_{\textrm{RF}}]_{1,\textrm{re}}} +\frac{\partial\mathcal{L}_{\textrm{bat}}}{\partial[\mathbf{r}]_{i,\textrm{im}}} \frac{\partial[\mathbf{r}]_{i,\textrm{im}}}{\partial[\bar{\mathbf{f}}_{\textrm{RF}}]_{1,\textrm{re}}}\right)\nonumber\\
&&+\!\sum_{j=1}^{N_{\textrm{R}}^{\textrm{RF}}}\!\left(\frac{\partial\mathcal{L}_{\textrm{bat}}}{\partial[\mathbf{H}_{\textrm{eq}}]_{j,1,\textrm{re}}} \frac{\partial[\mathbf{H}_{\textrm{eq}}]_{j,1,\textrm{re}}}{\partial[\bar{\mathbf{f}}_{\textrm{RF}}]_{1,\textrm{re}}} \!+\!\frac{\partial\mathcal{L}_{\textrm{bat}}}{\partial[\mathbf{H}_{\textrm{eq}}]_{j,1,\textrm{im}}} \frac{\partial[\mathbf{H}_{\textrm{eq}}]_{j,1,\textrm{im}}}{\partial[\bar{\mathbf{f}}_{\textrm{RF}}]_{1,\textrm{re}}}\right)\!,\nonumber\\
&&
\end{eqnarray}
\begin{eqnarray}
\label{eqn_derL_fRF_1im}
&&\frac{\partial\mathcal{L}_{\textrm{bat}}}{\partial[\bar{\mathbf{f}}_{\textrm{RF}}]_{1,\textrm{im}}} =\sum_{i=1}^{N_{\textrm{s}}}\left(\frac{\partial\mathcal{L}_{\textrm{bat}}}{\partial[\mathbf{r}]_{i,\textrm{re}}} \frac{\partial[\mathbf{r}]_{i,\textrm{re}}}{\partial[\bar{\mathbf{f}}_{\textrm{RF}}]_{1,\textrm{im}}} +\frac{\partial\mathcal{L}_{\textrm{bat}}}{\partial[\mathbf{r}]_{i,\textrm{im}}} \frac{\partial[\mathbf{r}]_{i,\textrm{im}}}{\partial[\bar{\mathbf{f}}_{\textrm{RF}}]_{1,\textrm{im}}}\right)\nonumber\\
&&+\!\sum_{j=1}^{N_{\textrm{R}}^{\textrm{RF}}}\!\left(\frac{\partial\mathcal{L}_{\textrm{bat}}}{\partial[\mathbf{H}_{\textrm{eq}}]_{j,1,\textrm{re}}} \frac{\partial[\mathbf{H}_{\textrm{eq}}]_{j,1,\textrm{re}}}{\partial[\bar{\mathbf{f}}_{\textrm{RF}}]_{1,\textrm{im}}} \!+\!\frac{\partial\mathcal{L}_{\textrm{bat}}}{\partial[\mathbf{H}_{\textrm{eq}}]_{j,1,\textrm{im}}} \frac{\partial[\mathbf{H}_{\textrm{eq}}]_{j,1,\textrm{im}}}{\partial[\bar{\mathbf{f}}_{\textrm{RF}}]_{1,\textrm{im}}}\right)\!.\nonumber\\
&&
\end{eqnarray}
By considering (\ref{eqn_fRF_reim})$-$(\ref{eqn_derL_fRF_1im}), we arrive at
\begin{eqnarray}
\label{eqn_derL_pha_1}
\frac{\partial\mathcal{L}_{\textrm{bat}}}{\partial[\boldsymbol{\phi}_{\textrm{P}}]_{1}} &&=\frac{\partial\mathcal{L}_{\textrm{bat}}}{\partial[\bar{\mathbf{f}}_{\textrm{RF}}]_{1,\textrm{re}}} \frac{\partial[\bar{\mathbf{f}}_{\textrm{RF}}]_{1,\textrm{re}}}{\partial[\boldsymbol{\phi}_{\textrm{P}}]_{1}} +\frac{\partial\mathcal{L}_{\textrm{bat}}}{\partial[\bar{\mathbf{f}}_{\textrm{RF}}]_{1,\textrm{im}}} \frac{\partial[\bar{\mathbf{f}}_{\textrm{RF}}]_{1,\textrm{im}}}{\partial[\boldsymbol{\phi}_{\textrm{P}}]_{1}}\nonumber\\
&&=-\sin[\boldsymbol{\phi}_{\textrm{P}}]_{1}\frac{\partial\mathcal{L}_{\textrm{bat}}}{\partial[\bar{\mathbf{f}}_{\textrm{RF}}]_{1,\textrm{re}}} +\cos[\boldsymbol{\phi}_{\textrm{P}}]_{1}\frac{\partial\mathcal{L}_{\textrm{bat}}}{\partial[\bar{\mathbf{f}}_{\textrm{RF}}]_{1,\textrm{im}}},
\end{eqnarray}

\emph{CP-NN:} The proof is similar to that of PP-NN and thus is omitted for simplicity.

Now we have shown $\mathcal{L}_{\textrm{bat}}$ is differentiable with respect to each parameter in $\boldsymbol\Theta$, which completes the proof.
\end{IEEEproof}


It can be seen that the proposed DL-JHPF is abstracted into an integrated DNN, where the hybrid processing matrices, $\mathbf{F}_{\textrm{RF}}$, $\mathbf{F}_{\textrm{BB}}$, $\mathbf{W}_{\textrm{RF}}$, and $\mathbf{W}_{\textrm{BB}}$, are essentially the trainable weights therein. From the proof of Theorem 1, each weight of this integrated DNN can be optimized iteratively through BP algorithm by minimizing the BCE loss. Therefore, the optimal precoding and combining matrices on training set are obtained.

For the NNs in Fig.~\ref{visual_model}, each dense layer is with rectified linear unit (ReLU) activation function and followed by a BN layer to avoid gradient diffusion and overfitting. The number of dense layers and the number of neurons in each dense layer need to be adjusted according to the input and output dimensions. Since the outputs of the NNs will be used for hybrid processing at the transmitter and the reciever, the activation functions of the output layers should be carefully designed and are elaborated as follows.

\emph{PP-NN and CP-NN:} The two NNs generate the phases for $\mathbf{F}_{\textrm{RF}}$ and $\mathbf{W}_{\textrm{RF}}$, respectively. Since (\ref{eqn_fRF}) and (\ref{eqn_wRF}) are periodic functions, ReLU activation function is used in the output layer to provide the unbiased output for all possible phases. We may also use Sigmoid or hyperbolic tangent as the activation function, after which the outputs are multiplied by $2\pi$ or $\pi$ to obtain the final phases with the range of $[0,2\pi]$ or $[-\pi,\pi]$. According to the simulation trails, ReLU and hyperbolic tangent achieve almost the same performance while Sigmoid performs worse. Therefore, ReLU is preferable since it is simple and free of the operation of exponential functions.

\emph{Re/ImDP-NN and Re/ImDC-NN:} The four NNs generate the real and imaginary parts for $\mathbf{F}_{\textrm{BB}}$ and $\mathbf{W}_{\textrm{BB}}$, respectively. Since $\mathbf{F}_{\textrm{BB}}$ can be normalized by (\ref{eqn_DP_norm}) while $\mathbf{W}_{\textrm{BB}}$ has no constraint, the output layers do not apply any activation function to impose constraints and directly output the values that are input into the neurons.

\emph{NN demodulator:} This NN approximates the original bits, $\mathbf{X}_{\textrm{b}}$, based on $\mathbf{r}$. The approximation for each element in $\mathbf{X}_{\textrm{b}}$ is a binary classification and thus the Sigmoid activation function is used for the output layer of the NN demodulator.

\vspace{-0.2cm}
\subsection{Deployment and Testing}

In this subsection, we elaborate the deployment and testing of the trained DL-JHPF for practical implementation, where $\mathbf{H}$ is assumed to be available at both the transmitter and the receiver.\footnote{Although only the estimated channel is available in practical implementation, it has been shown by the simulation results that the relatively accurate channel estimate hardly causes performance loss and is almost equivalent to $\mathbf{H}$.}

The practical deployment of DL-JHPF includes the following three parts:

\emph{Deployment of hybrid processing designer:} PP-NN and CP-NN will be deployed together at \emph{both the transmitter and the receiver} to output the analog processing matrices, $\mathbf{F}_{\textrm{RF}}$ and $\mathbf{W}_{\textrm{RF}}$, based on which the equivalent channel, $\mathbf{H}_{\textrm{eq}}$, can be generated via (\ref{eqn_Heq}). ReDP-NN and ImDP-NN are equipped at the \emph{transmitter} to generate the digital precoder, $\mathbf{F}_{\textrm{BB}}$, while ReDC-NN and ImDC-NN are equipped at the \emph{receiver} to generate the digital combiner, $\mathbf{W}_{\textrm{BB}}$, both based on $\mathbf{H}_{\textrm{eq}}$.

\emph{Deployment of signal flow simulator:} It is only used for the training stage and will be replaced by the actual transceiver and wireless fading channel in the deployment and testing stage.

\emph{Deployment of NN demodulator:} It will be deployed at the \emph{receiver} to output the recovered bits, $\hat{\mathbf{X}}_{\textrm{b}}$, based on the detected signal, $\mathbf{r}$, after compensating the impact of the fading channel.

When testing the trained DL-JHPF in real world, the channel may change rapidly due to the relative motion of the transceiver and scatterers, in which case DL-JHPF will be faced new propagation scenarios with different channel statistics from the training stage. This channel scenario discrepancy poses a high requirement on the robustness of DL-JHPF. Fortunately, the offline trained framework in Section III.C is quite robust to the new channel scenarios that are not observed before as shown from our simulation results (Figs.~\ref{HBD_BS_proDL_UMaipCSI} and \ref{HBD_BS_proDL_ofdm_UMaipCSI}). The further online fine-tuning may only provide marginal performance improvement but requires a relatively large overhead and needs to be performed frequently in the rapidly changed channel scenario. In addition, only the NNs at the receiver can be fine-tuned and thus the performance after fine-tuning will still have an intrinsic loss compared to the end-to-end training in Section III.C. To sum up, the proposed framework can cope with the mismatch of the channel scenario without relying on the fine-tuning in most cases.

\subsection{Complexity Analysis}

In this subsection, we analyze the computational complexity of the proposed DL-JHPF in testing stage by using the metric of required number of floating point operations (FLOPs). According to Fig.~\ref{visual_model}, the total required FLOPs of all neural layers in DL-JHPF is given by
\begin{eqnarray}
\label{eqn_Complexity_NN}
\mathcal{C}_{\textrm{NN}}\sim\mathcal{O}\left(\sum_{\Delta\in\mathcal{N}}\sum_{i=1}^{L^{\Delta}-1}N^{\Delta}_{i}N^{\Delta}_{i+1}\right),
\end{eqnarray}
where $\mathcal{N}$ denotes the set including all NNs in DL-JHPF, $L^{\Delta}$ and $N^{\Delta}_{i}$ represent the number of neural layers and the number of neurons of the $i$th neural layer of the NN $\Delta$.

In addition, the complexity of matrix multiplications in the framework is given by
\begin{eqnarray}
\label{eqn_Complexity_matrix}
\mathcal{C}_{\textrm{Mat}}\sim\mathcal{O}\bigl(N_{\textrm{R}}^{\textrm{RF}}N_{\textrm{T}}N_{\textrm{R}}).
\end{eqnarray}

Then, the total complexity of the proposed DL-JHPF can be expressed as
\begin{eqnarray}
\label{eqn_Complexity_DL_JHPF}
\mathcal{C}_{\textrm{DL-JHPF}}\sim\mathcal{C}_{\textrm{NN}}+\mathcal{C}_{\textrm{Mat}}.
\end{eqnarray}
It is noted that the NNs can be run efficiently via parallel computing on the graphic processing unit (GPU) and the simple matrix multiplications only cause negligible computational load for the central processing unit (CPU) compared with the existing schemes. Therefore, the proposed DL-JHPF is with low complexity and consumes the very limited runtime.

\subsection{Extension to OFDM Systems}

In this subsection, we extend the proposed DL-JHPF to the wideband OFDM systems. Two key issues need to be considered for the extension:
\begin{itemize}[\IEEEsetlabelwidth{Z}]
\item[1)] In the OFDM systems, the digital precoder and combiner can be designed independently for different subcarriers while the analog precoder and combiner must be shared by all subcarriers. It is critical to design the unified analog precoder and combiner performing well for all subcarriers.

\item[2)] It is important to maintain the relatively small size, i.e., the number of hidden layers and the number of neurons in each layer in the NNs, and short training time for DL-JHPF when the number of subcarriers is large.
\end{itemize}

In the following, we study how to address the two issues when extending DL-JHPF to the OFDM systems.

According to \cite{P. Dong_b}, the $N_{\textrm{R}}\times N_{\textrm{T}}$ channel matrix between the receiver and the transmitter of the $k$th subcarrier can be expressed as
\begin{eqnarray}
\label{eqn_Hk}
\mathbf{H}[k]&&=\beta\sum_{n=1}^{N_{\textrm{cl}}}\sum_{m=1}^{N_{\textrm{ray}}} \alpha_{n,m}e^{-j2\pi\tau_n f_s\frac{k}{K}}\mathbf{a}_{\textrm{R}}(\varphi_{n,m})\mathbf{a}_{\textrm{T}}^{H}(\phi_{n,m}),
\end{eqnarray}
where $\beta=\sqrt{\frac{N_{\textrm{T}}N_{\textrm{R}}}{N_{\textrm{cl}}N_{\textrm{ray}}}}$, $\tau_n$, $f_s$, and $K$ denote the delay of the $n$th cluster, the sampling rate, and the number of OFDM subcarriers, respectively.
The signal transmission model in (\ref{eqn_r}) becomes subcarrier dependent and the detected signal of the $k$th subcarrier is given by\footnote{Although $\mathbf{x}$ and $\mathbf{n}$ are also different for different subcarriers, they are independent of the channel and thus the index $k$ in them is omitted.}
\setlength{\arraycolsep}{0.05em}
\begin{eqnarray}
\label{eqn_rk}
\mathbf{r}[k]&&=\sqrt{P}\mathbf{W}_{\textrm{BB}}^H[k]\mathbf{W}_{\textrm{RF}}^H\mathbf{H}[k]\mathbf{F}_{\textrm{RF}}\mathbf{F}_{\textrm{BB}}[k]\mathbf{x}\nonumber\\
&&\quad+\mathbf{W}_{\textrm{BB}}^H[k]\mathbf{W}_{\textrm{RF}}^H\mathbf{n}.
\end{eqnarray}

In the following, we propose a simple method to design the structure of training data so that the DL-JHPF in Section III.C can be flexibly extended to OFDM systems without changing the framework architecture. That is, both the framework size and training time will not be increased. The process of training and testing is detailed as follows.

\begin{figure}[t]
\centering
\includegraphics[width=3.5in]{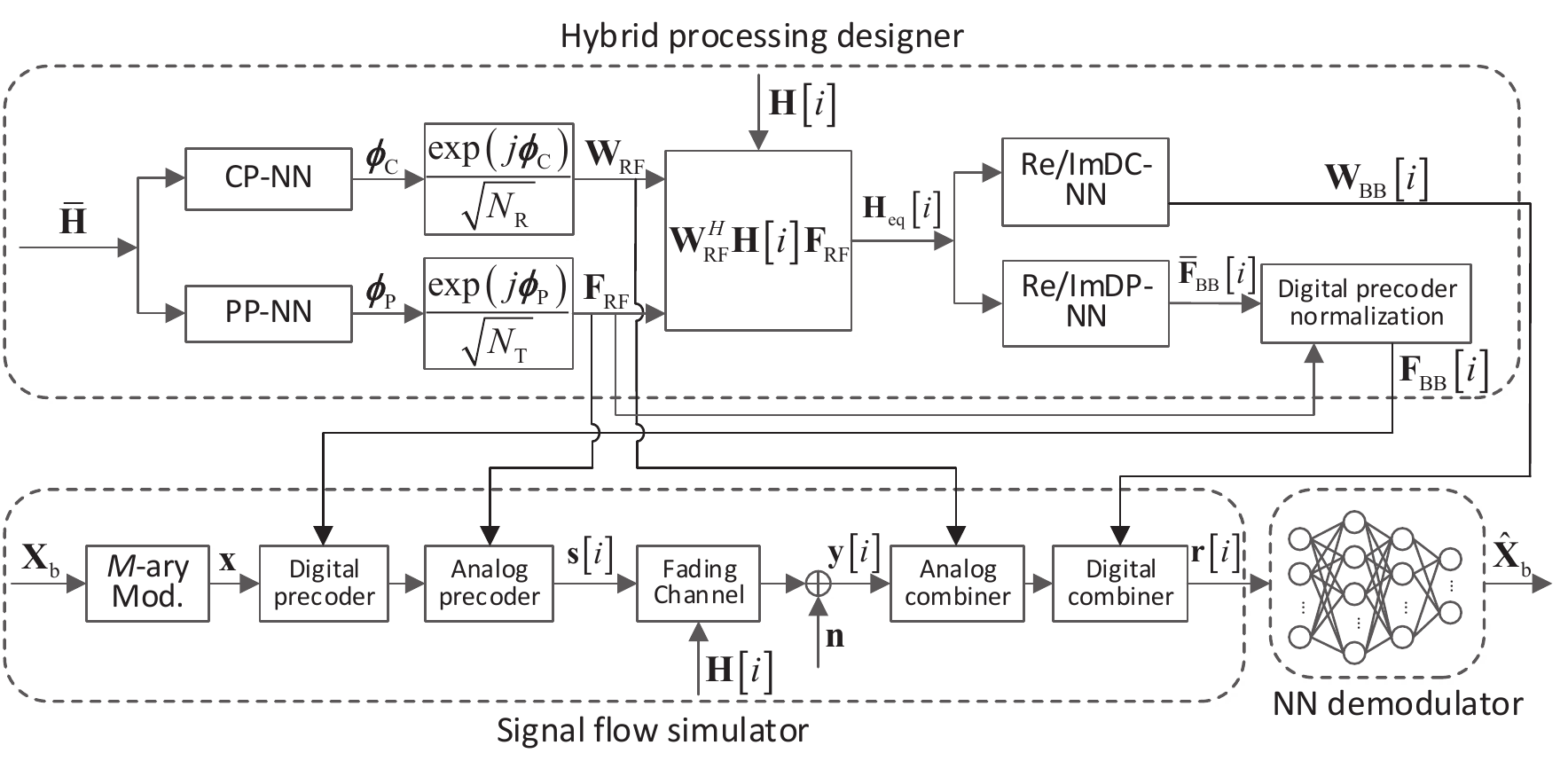}
\caption{Extension of the proposed DL-JHPF to OFDM systems.}\label{DL_framework_ofdm}
\end{figure}

\emph{Training:} Compared to the training sample with the input tuple $\langle\mathbf{H}, \mathbf{X}_{\textrm{b}}, \mathbf{n}\rangle$ in Section III.C, we modify the input tuple as $\langle\bar{\mathbf{H}}, \mathbf{H}[i], \mathbf{X}_{\textrm{b}}, \mathbf{n}\rangle$, where $\bar{\mathbf{H}}$ is the channel matrix of a given subcarrier, e.g., the $q$th subcarrier, same for all training samples while $\mathbf{H}[i]$ is the channel matrix of an uncertain subcarrier with $i$ randomly generated from the set $\{1,2,\ldots,K\}$ for each training sample. As shown in Fig.~\ref{DL_framework_ofdm}, when inputting each training sample into the framework, $\bar{\mathbf{H}}$ will be used to generate $\mathbf{F}_{\textrm{RF}}$ and $\mathbf{W}_{\textrm{RF}}$ via PP-NN and CP-NN. Then $\mathbf{F}_{\textrm{RF}}$ and $\mathbf{W}_{\textrm{RF}}$ along with $\mathbf{H}[i]$ are used to generate the equivalent channel of the $i$th subcarrier, $\mathbf{H}_{\textrm{eq}}[i]$, based on which, $\mathbf{F}_{\textrm{BB}}[i]$ and $\mathbf{W}_{\textrm{BB}}[i]$ can be obtained through Re/ImDP-NN and Re/ImDC-NN. On the other hand, $\mathbf{H}[i]$ is also input into the signal flow simulator to act as the fading channel since this training sample is used to simulate the transmission of the $i$th subcarrier. Then the end-to-end training can be performed by minimizing the BCE loss between $\mathbf{X}_{\textrm{b}}$ and $\hat{\mathbf{X}}_{\textrm{b}}$. Through training, we can obtain the unified analog precoder and combiner that match the channel of each subcarrier well without complicating the architecture of DL-JHPF.

\emph{Testing:} With $\mathbf{H}[k], k=1,2,\ldots,K$, available at the transceiver, choose the channel matrix of the $q$th subcarrier as $\bar{\mathbf{H}}$. Input $\bar{\mathbf{H}}$ into PP-NN and CP-NN to generate the unified $\mathbf{F}_{\textrm{RF}}$ and $\mathbf{W}_{\textrm{RF}}$ for all subcarriers. The unified $\mathbf{F}_{\textrm{RF}}$ and $\mathbf{W}_{\textrm{RF}}$ along with the channel of each subcarrier, $\mathbf{H}[k], k=1,2,\ldots,K$, are used to generate the corresponding equivalent channel, which will be input into Re/ImDP-NN and Re/ImDC-NN to generate $\mathbf{F}_{\textrm{BB}}[k]$ and $\mathbf{W}_{\textrm{BB}}[k]$ for channel equalization in each subcarrier. The NN demodulator will be used to recover the original bits for each subcarrier based on the detected signal, $\mathbf{r}[k]$.

\section{Simulation Results}

In this section, the effectiveness of the proposed DL-JHPF is verified in several cases. Six hybrid processing schemes and the fully-digital transceiver architecture are used as the baseline schemes for comparison: 1) HBD scheme in \cite{W. Ni}; 2) Beam sweeping (BeS) scheme in \cite{O. E. Ayach}; 3) Discrete Fourier transform (DFT) codebook based joint digital beamforming (DCJDB) scheme, where the analog precoder and combiner are searched from the DFT codebook by the method in \cite{A. M. Elbir} while the digital precoder and combiner are jointly optimized according to \cite{D. P. Palomar}; 4) Joint digital beamforming with alternating minimization (JDB-AltMin), where the optimal precoding and combining matrices are first designed according to \cite{D. P. Palomar}, based on which the hybrid precoding and combining matrices are constructed according to the PE-AltMin algorithm in \cite{X. Yu}; 5) Hybrid beamforming via deep learning (HBDL) scheme in \cite{A. M. Elbir}; 6) Deep learning for direct hybrid precoding (DLDHP) scheme in \cite{X. Li}; 7) Fully-digital transceiver architecture.

\subsection{Simulation Settings}

\begin{table}[!t]
\centering
\caption{Architectures of DNNs in Proposed DL-HPF}
\label{table_1}
\begin{tabular}{p{1.4cm}<{\centering}|c|c|c}
\hline
~ & Layer type & \makecell{Number of \\neurons} & \makecell{Activation \\function}\\
\hline
\multirow{5}{*}{PP-NN} & Input & $2N_{\textrm{T}}N_{\textrm{R}}$ & - \\
\cline{2-4}
~ & Dense & 512 & ReLU\\
\cline{2-4}
~ & Dense & 256 & ReLU\\
\cline{2-4}
~ & Dense & 128 & ReLU\\
\cline{2-4}
~ & Output & $N_{\textrm{T}}N_{\textrm{T}}^{\textrm{RF}}$ & ReLU\\
\hline
\multirow{6}{*}{CP-NN} & Input & $2N_{\textrm{T}}N_{\textrm{R}}$ & - \\
\cline{2-4}
~ & Dense & 512 & ReLU\\
\cline{2-4}
~ & Dense & 256 & ReLU\\
\cline{2-4}
~ & Dense & 128 & ReLU\\
\cline{2-4}
~ & Dense & 64 & ReLU\\
\cline{2-4}
~ & Output & $N_{\textrm{R}}N_{\textrm{R}}^{\textrm{RF}}$ & ReLU\\
\hline
\multirow{5}{*}{\makecell{Re/ImDP-\\NN}} & Input & $2N_{\textrm{T}}^{\textrm{RF}}N_{\textrm{R}}^{\textrm{RF}}$ & - \\
\cline{2-4}
~ & Dense & 20 & ReLU\\
\cline{2-4}
~ & Dense & 40 & ReLU\\
\cline{2-4}
~ & Dense & 20 & ReLU\\
\cline{2-4}
~ & Output & $N_{\textrm{T}}^{\textrm{RF}}N_{\textrm{s}}$ & - \\
\hline
\multirow{5}{*}{\makecell{Re/ImDC-\\NN}} & Input & $2N_{\textrm{T}}^{\textrm{RF}}N_{\textrm{R}}^{\textrm{RF}}$ & - \\
\cline{2-4}
~ & Dense & 20 & ReLU\\
\cline{2-4}
~ & Dense & 40 & ReLU\\
\cline{2-4}
~ & Dense & 20 & ReLU\\
\cline{2-4}
~ & Output & $N_{\textrm{R}}^{\textrm{RF}}N_{\textrm{s}}$ & - \\
\hline
\multirow{5}{*}{\makecell{NN\\demodulator}} & Input & $2N_{\textrm{s}}$ & - \\
\cline{2-4}
~ & Dense & 20 & ReLU\\
\cline{2-4}
~ & Dense & 50 & ReLU\\
\cline{2-4}
~ & Dense & 20 & ReLU\\
\cline{2-4}
~ & Output & $N_{\textrm{s}}\log_{2} M$ & Sigmoid \\
\hline
\end{tabular}
\vspace{-0.3cm}
\end{table}

\emph{1) System Settings:} We set $N_{\textrm{T}}=32$ and $N_{\textrm{T}}^{\textrm{RF}}=3$ for the transmitter and $N_{\textrm{R}}=16$ and $N_{\textrm{R}}^{\textrm{RF}}=3$ for the receiver. The number of data streams is set as $N_{\textrm{s}}=3$. The channel data are generated according to the 3GPP TR 38.901 Release 15 channel model \cite{3GPP}. Specifically, we use the clustered delay line models with $N_{\textrm{cl}}=3$ clusters and $N_{\textrm{ray}}=20$ rays in each cluster. The carrier frequency is $f_c=28$ GHz. For OFDM systems, the sampling rate is $f_s=100$ MHz and the number of subcarriers is $K=64$. Two channel scenarios, urban micro (UMi) street non-line of sight (NLOS) scenario and urban macro (UMa) NLOS scenario, are considered.\footnote{According to the parameters for UMi NLOS scenario and UMa NLOS scenario defined by \cite{3GPP}, we use the system object, nr5gCDLChannel, embedded in 5G Library for LTE System Toolbox in MATLAB to generate the corresponding channel data.} Quadrature phase shift keying (QPSK) is used as the modulation method.

\emph{2) Proposed DL-JHPF Settings:} The training set, validation set, and testing set contain $261$,$000$, $29$,$000$, and $10$,$000$ samples, respectively. The training set and validation set are generated in UMi NLOS scenario while the testing set is generated in both UMi NLOS and UMa NLOS scenarios. Adam is used as the optimizer. The number of epochs in the training stage is set as $800$ while the corresponding learning rates are $10^{-3}$ for the first $500$ epochs and $10^{-4}$ for the rest $300$ epochs, respectively. The batch size is $256$. The architecture of each NN in DL-JHPF is listed in Table~\ref{table_1}, where the BN layer is added after each dense layer and thus is not listed in the table for simplicity.

\subsection{Performance Evaluation}

\begin{figure}[!t]
\centering
\includegraphics[trim=0 0 0 0, width=3.6in]{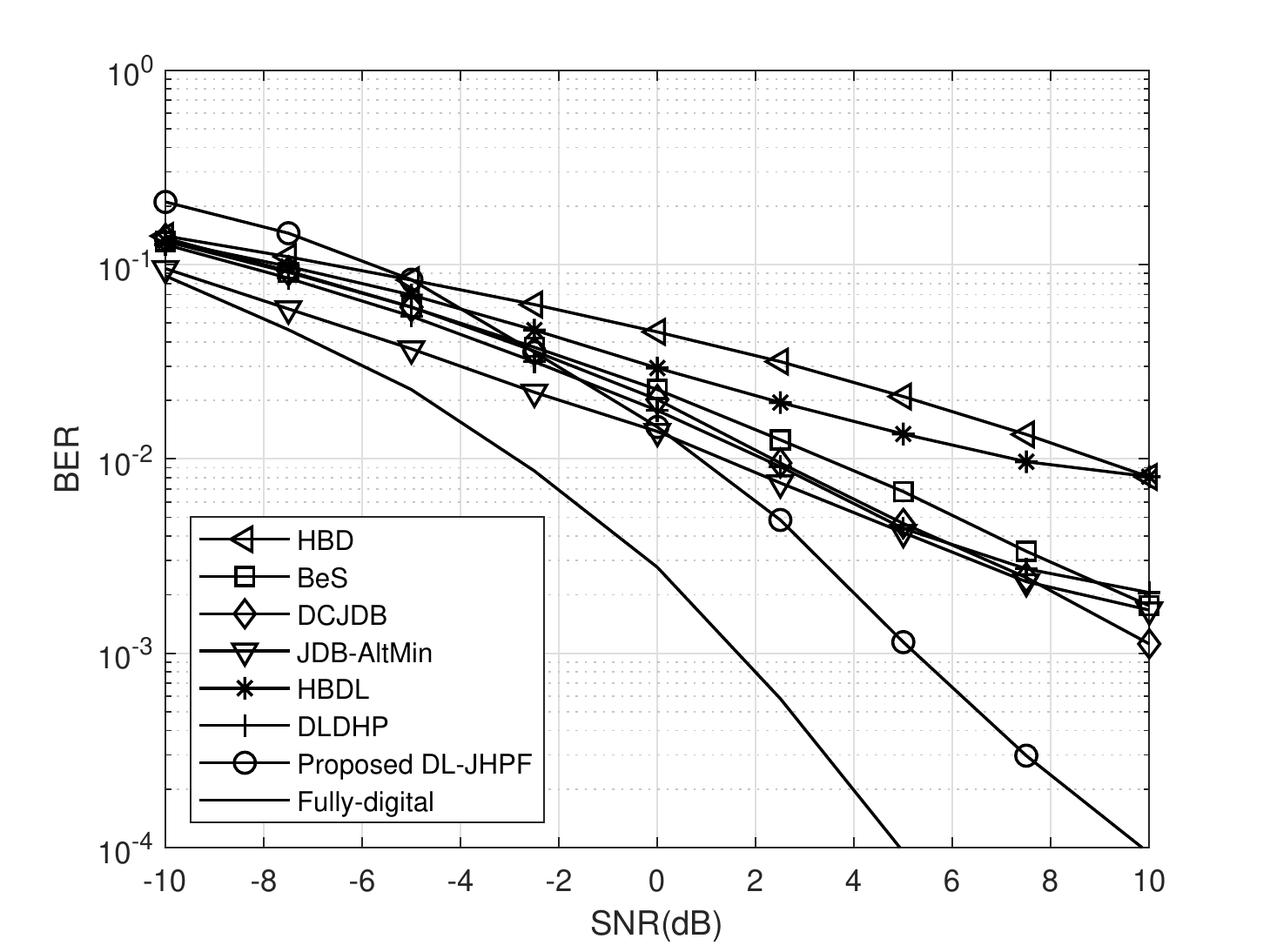}
\caption{BER performance of the proposed DL-JHPF and the existing hybrid processing schemes.}\label{HBD_BS_proDL}
\end{figure}

\begin{figure}[!t]
\centering
\includegraphics[trim=0 0 0 0, width=3.6in]{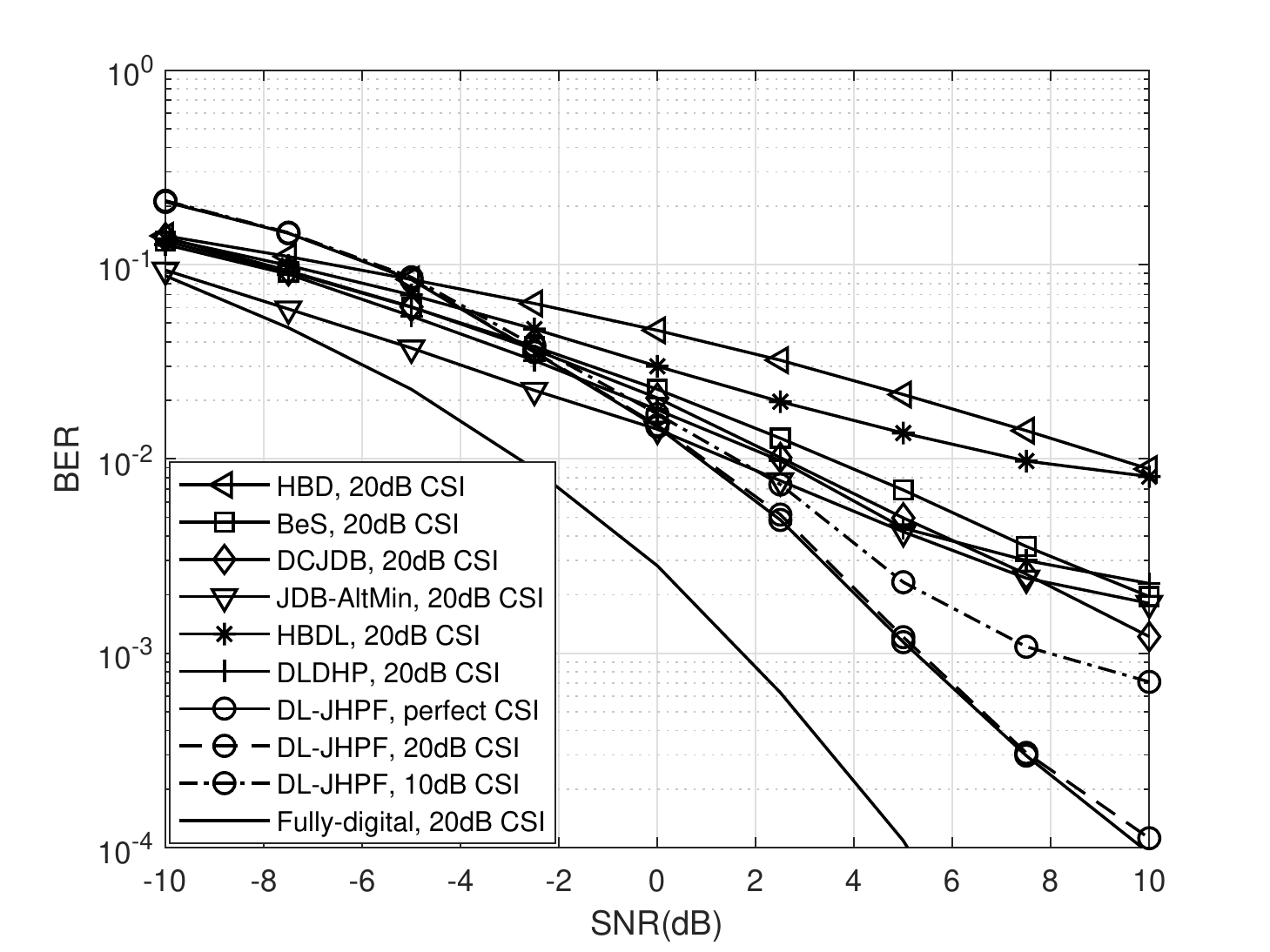}
\caption{Robustness of the proposed DL-JHPF with mismatched CSI.}\label{HBD_BS_proDL_ipCSI}
\end{figure}

\begin{figure}[!t]
\centering
\includegraphics[trim=0 0 0 0, width=3.6in]{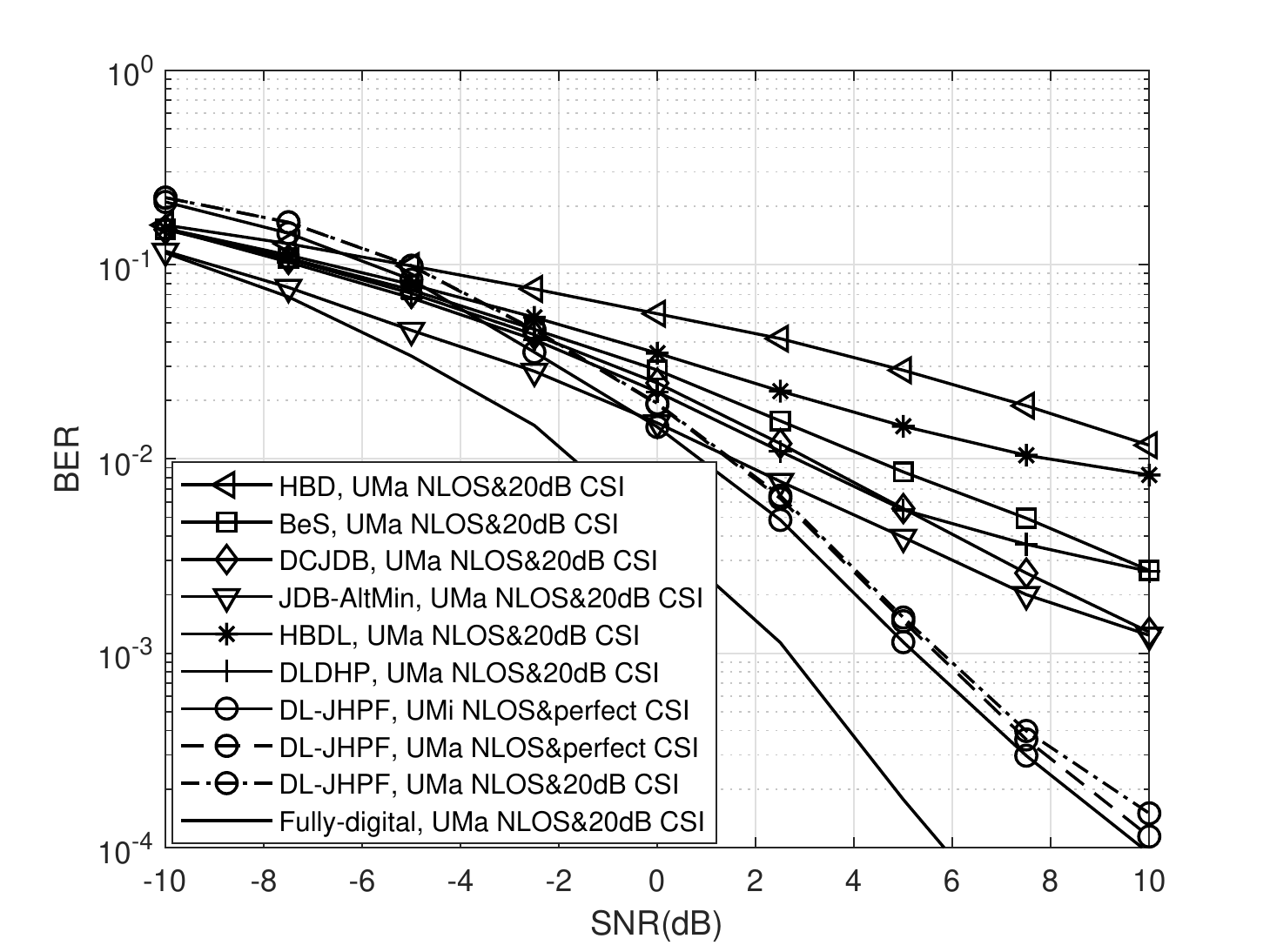}
\caption{Robustness of the proposed DL-JHPF with mismatched channel scenario and CSI.}\label{HBD_BS_proDL_UMaipCSI}
\end{figure}

\begin{figure}[!t]
\centering
\includegraphics[trim=0 0 0 0, width=3.6in]{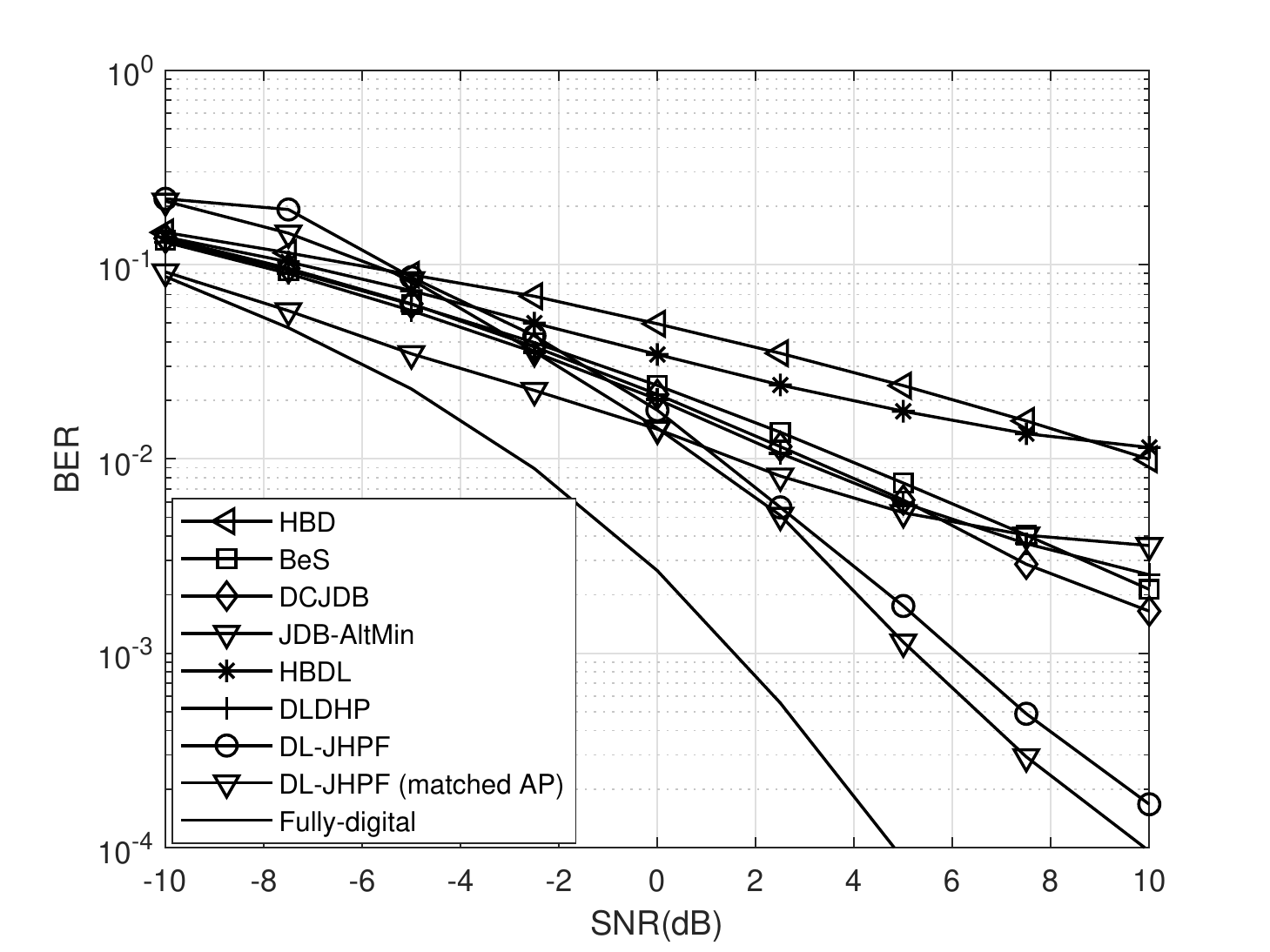}
\caption{BER performance of the proposed DL-JHPF and the existing hybrid processing schemes in OFDM systems.}\label{HBD_BS_proDL_ofdm}
\end{figure}

\begin{figure}[!t]
\centering
\includegraphics[trim=0 0 0 0, width=3.6in]{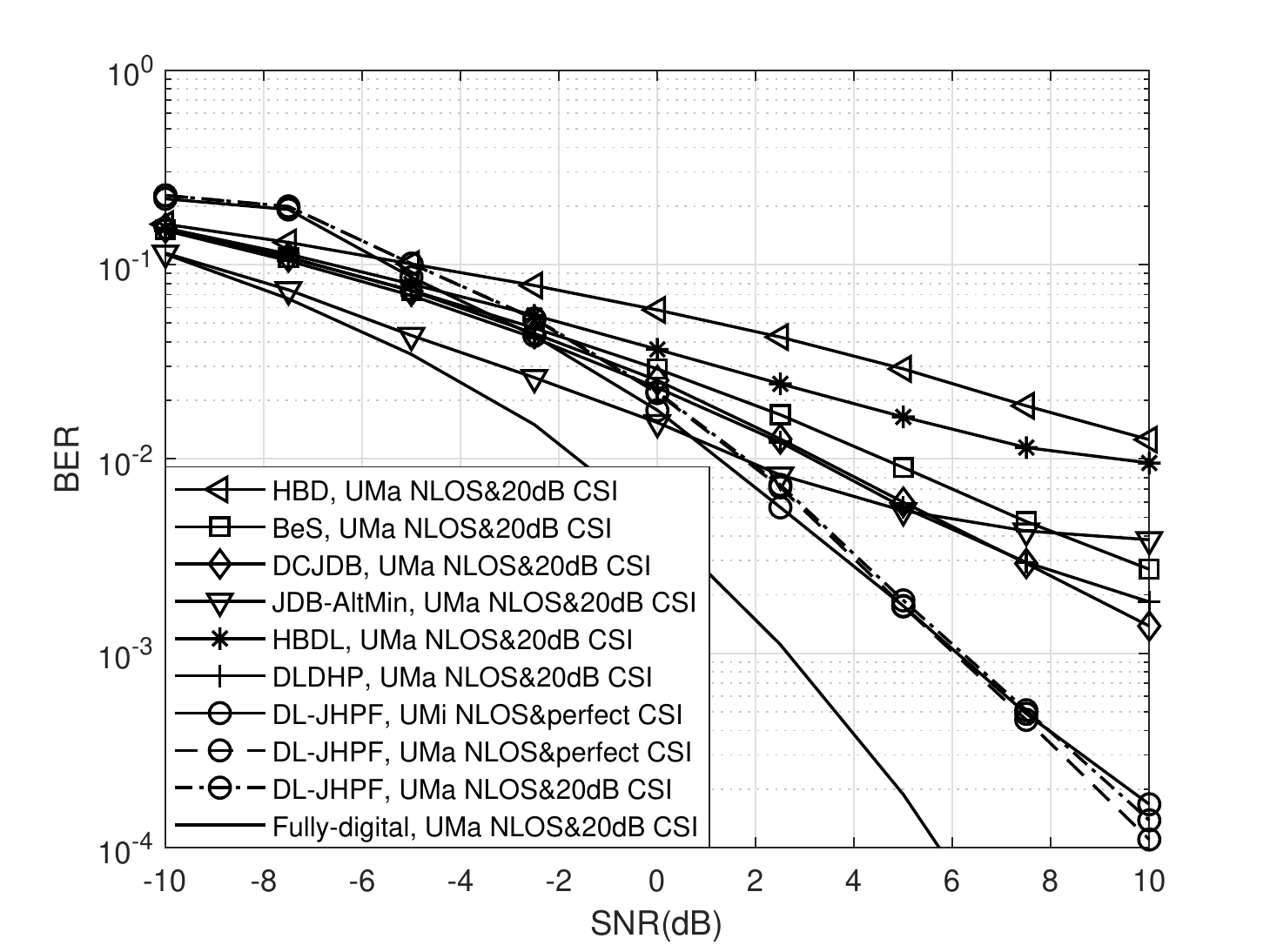}
\caption{Robustness of the proposed DL-JHPF with mismatched channel scenario and CSI in OFDM systems.}\label{HBD_BS_proDL_ofdm_UMaipCSI}
\end{figure}

In Figs.~\ref{HBD_BS_proDL}$-$\ref{HBD_BS_proDL_UMaipCSI}, the proposed DL-JHPF is first evaluated in narrowband systems while the performance in wideband OFDM systems is presented in Figs.~\ref{HBD_BS_proDL_ofdm} and \ref{HBD_BS_proDL_ofdm_UMaipCSI}.

Fig.~\ref{HBD_BS_proDL} shows the BER performance of HBD, BeS, DCJDB, JDB-AltMin, HBDL, DLDHP, the proposed DL-JHPF, and the fully-digital architecture versus signal-to-noise ratio (SNR) in UMi NLOS scenario with perfect CSI. From the figure, DL-JHPF has a larger slope for the BER curve and outperforms the other six hybrid processing schemes after $\textrm{SNR}=0$ dB although it performs not very well in the low SNR regime. When $\textrm{BER}=10^{-2}$, the proposed DL-JHPF achieves about $0.2$ dB, $1$ dB, $1.2$ dB, $2$ dB, $6$ dB, and $8$ dB gains compared to JDB-AltMin, DLDHP, DCJDB, BeS, HBDL, and HBD, respectively. The advantage of DL-JHPF becomes more obvious as SNR increases and the BER is smaller than $10^{-4}$ when $\textrm{SNR}=10$ dB while the performance of other four schemes is larger than $10^{-3}$. With the significantly increased number of RF chains, the fully-digital beamforming obtains substantial diversity gains, which directly leads to the better BER performance than all the hybrid processing schemes. The performance gap between the proposed DL-JHPF and the fully-digital beamforming is about 4dB.

Perfect CSI is used in framework training while only estimated CSI is available in the practical transmitter and receiver, which leads to the CSI mismatch. In Fig.~\ref{HBD_BS_proDL_ipCSI}, we investigate the robustness of the proposed DL-JHPF with mismatched CSI, where the BER curve tested with perfect CSI in Fig.~\ref{HBD_BS_proDL} is also plotted as the lower bound. We use the approach in \cite{P. Dong_b} to estimate channels at $\textrm{SNR}=10$ dB and $20$ dB, respectively, for hybrid processing design. From Fig.~\ref{HBD_BS_proDL_ipCSI}, when tested with the CSI estimated at $20$ dB, DL-JHPF achieves almost the same BER performance as the perfect CSI case and outperforms the other six hybrid processing schemes after $\textrm{SNR}=0$ dB, indicating that DL-JHPF is hardly impacted by the mismatched CSI estimated at $20$ dB. When tested with the CSI estimated at $10$ dB, performance loss occurs at an acceptable level for DL-JHPF. The loss is less than $1$ dB when $\textrm{BER}=10^{-2}$ and DL-JHPF still has the clear performance superiority after $\textrm{SNR}=2.5$ dB even compared to other hybrid processing schemes with the CSI estimated at $20$ dB.

As mentioned in Section III.D, it is very likely to face with different channel scenarios in the practical testing for DL-JHPF. In Fig.~\ref{HBD_BS_proDL_UMaipCSI}, we further consider this channel scenario mismatch and test the robustness of DL-JHPF to the aggregate impact caused by channel scenario and CSI mismatch. For DL-JHPF, the BER performance is tested in UMi NLOS scenario with perfect CSI, in UMa NLOS scenario with perfect CSI (mismatched channel scenario), and in UMa NLOS scenario with the CSI estimated at $20$ dB (mismatched channel scenario and CSI), respectively. The performance curves of the baseline schemes evaluated in UMa NLOS scenario with the CSI estimated at $20$ dB are also plotted for comparison. From Fig.~\ref{HBD_BS_proDL_UMaipCSI}, the channel scenario mismatch causes only less than $0.5$ dB performance loss for DL-JHPF. The total loss caused by the aggregate impact of channel scenario and CSI mismatch is only less than $1$ dB. The proposed DL-JHPF has learned the inherent structure of the mmWave channels and thus is able to maintain its advantage even with mismatched channel scenarios and CSI.

Fig.~\ref{HBD_BS_proDL_ofdm} shows the BER performance of HBD, BeS, DCJDB, JDB-AltMin, HBDL, DLDHP, the proposed DL-JHPF, and the fully-digital architecture in OFDM systems with UMi NLOS scenario and perfect CSI, which is similar to that in Fig.~\ref{HBD_BS_proDL}. In addition, we plot the BER performance of an ideal case with matched analog processing (AP) for DL-JHPF, where different analog processing matrices are designed for different subcarriers to match the corresponding channels. This is impossible to be implemented in practical systems and we just use it to quantify the performance loss caused by using the unified analog processing matrices for all subcarriers. From Fig.~\ref{HBD_BS_proDL_ofdm}, only about $1$ dB loss is incurred, which proves the effectiveness of DL-JHPF in OFDM systems by simply modifying the structure of training data without changing the framework architecture and increasing the training time.

In Fig.~\ref{HBD_BS_proDL_ofdm_UMaipCSI}, we further test the robustness of DL-JHPF in OFDM systems with the mismatched channel scenario and CSI. The aggregate impact of channel scenario and CSI mismatch is still limited, and DL-JHPF tested in UMa NLOS scenario with the CSI estimated at $20$ dB (mismatched channel scenario and CSI) even outperforms that tested in UMi NLOS scenario with perfect CSI after $\textrm{SNR}=8$ dB, which verifies the effectiveness and robustness of the proposed DL-JHPF in OFDM systems. In addition, the performance gap between the proposed DL-JHPF and the fully-digital beamforming maintains at about 4dB.

\subsection{Computational Complexity Comparison}

\begin{table}
  \centering
  \caption{Runtime of Hybrid Processing Schemes}
  \label{runtime}
  \begin{tabular}{c|c}
  \hline
  ~ & Runtime (in ms)\\
  \hline
  HBD & $6.98$\\
  \hline
  BeS & $10.61$\\
  \hline
  DCJDB & $338.73$\\
  \hline
  JDB-AltMin & $1.56$\\
  \hline
  HBDL & $0.46$\\
  \hline
  DLDHP & $0.16$\\
  \hline
  Proposed DL-JHPF & $0.06$\\
  \hline
  \end{tabular}
\end{table}

For mmWave mobile communications, the length of coherence time becomes smaller compared to that in sub-6 GHz and thus the runtime of a hybrid processing scheme is vital. Based on the simulation settings mentioned above, we compare the runtime of the proposed DL-JHPF in the testing stage with the baseline schemes in Table~\ref{runtime}. The HBD, BeS, DCJDB, and JDB-AltMin schemes are run on the Intel(R) Core(TM) i7-3770 CPU while the proposed DL-JHPF are run on the NVIDIA GeForce GTX 2080 Ti GPU. For HBDL and DLDHP, the predictions of $\mathbf{F}_{\textrm{RF}}$ and $\mathbf{W}_{\textrm{RF}}$ are implemented via DNN on the GPU while the following design of $\mathbf{F}_{\textrm{BB}}$ and $\mathbf{W}_{\textrm{BB}}$ are executed on the CPU. By moving the time-consuming design of analog processing to the GPU that enables the efficient parallel computing, the DL based schemes reduce the runtime significantly compared to the conventional schemes. Through carefully design, the proposed DL-JHPF is fully GPU-driven when generating hybrid processing matrices and thus consumes the minimum time among the three DL based schemes. Therefore, the proposed DL-JHPF is more suitable for mmWave communications, especially for the high-mobility scenario.

\section{Conclusion}

In this paper, DL is applied for joint hybrid processing design at the transceiver in mmWave massive MIMO systems. A novel DL-JHPF is developed to learn the optimal analog and digital processing matrices by minimizing the end-to-end BCE loss between the original and recovered bits. The elaborate architecture of the proposed DL-JHPF guarantees the BP-enabled training of each NN therein. By simply modifying the structure of training data, DL-JHPF can be flexibly extended to OFDM systems without changing the framework architecture and increasing the training time. Simulation results show the superiority and robustness of DL-JHPF in various non-ideal conditions with the significantly reduced runtime.

%



\ifCLASSOPTIONcaptionsoff
  \newpage
\fi


\end{document}